\documentclass[preprint]{aastex}
\usepackage{epsfig}
\usepackage{graphicx}
\begin{document}

\title{The Identification of Extreme Asymptotic Giant Branch Stars and Red Supergiants in M33 by 24\micron\ Variability}
\author{Edward J. Montiel\altaffilmark{1}, Sundar Srinivasan\altaffilmark{2}, Geoffrey C. Clayton\altaffilmark{1}, Charles W. Engelbracht\altaffilmark{3,4}, Christopher B. Johnson\altaffilmark{1}}
\altaffiltext{1}{Department of Physics \& Astronomy, Louisiana State University, Baton Rouge, LA 70803; emonti2@lsu.edu}
\altaffiltext{2}{Academica Sinica, Institute of Astronomy and Astrophysics, PO Box 23-141, Taipei 10617, Taiwan, R. O. C.}
\altaffiltext{3}{Steward Observatory, 933 North Cherry Avenue, Tucson, AZ 85721}
\altaffiltext{4}{Raytheon Company, 1151 E. Hermans Road, Tucson, AZ 85756, USA; Deceased}

\begin{abstract}
We present the first detection of 24~\micron\ variability in 24 sources in the Local Group galaxy M33. These results are based on 4 epochs of MIPS observations, which are irregularly spaced over $\sim$750 days. We find that these sources are constrained exclusively to the Holmberg radius of the galaxy, which increases their chances of being members of M33. We have constructed spectral energy distributions (SEDs) ranging from the optical to the sub-mm to investigate the nature of these objects. We find that 23 of our objects are most likely heavily self-obscured, evolved stars; while the remaining source is the Giant HII region, NGC 604. We believe that the observed variability is the intrinsic variability of the central star reprocessed through their circumstellar dust shells. Radiative transfer modeling was carried out to determine their likely chemical composition, luminosity, and dust production rate (DPR). As a sample, our modeling has determined an average luminosity of $(3.8 \pm 0.9) \times 10^4$ L$_\odot$ and a total DPR of $(2.3 \pm 0.1) \times 10^{-5}$ M$_\odot$ yr$^{-1}$. Most of the sources,  given the high DPRs and short wavelength obscuration, are likely ÒextremeÓ AGB (XAGB) stars. Five of the sources are found to have luminosities above the classical AGB limit (M$_{\rm bol} <$ --7.1 mag, L $>$ 54,000 L$_\odot$), which classifies them as probably red supergiants (RSGs). Almost all of the sources are classified as oxygen rich. As also seen in the LMC, a significant fraction of the dust in M33 is produced by a handful of XAGB and RSG stars.
\end{abstract}


\section{Introduction} 
The asymptotic giant branch (AGB) denotes the final phase of nuclear activity in low to intermediate mass stars. This advanced stage of stellar evolution is marked by cooler temperatures and higher luminosities as the outer envelope of the star swells from hundreds to thousands of times its initial main sequence size. At the start of the AGB phase nearly all of the stars are predominantly oxygen-rich (O-rich). As the stars continue to evolve and climb the AGB thermal pulses (TPs) begin, which allow for a process known as the third dredge-up to take place. This internal mixing brings built-up reserves of carbon from deeper in the star to the surface changing the composition from O-rich to carbon-rich (C-rich) after enough TPs have occurred. However, in the case of more massive AGB stars (M $\ga$ 4.5 M$_\odot$, Boothroyd et al. 1995) an additional event known as hot bottom burning (HBB, Sackmann \& Boothroyd 1992) occurs, which converts the C-rich photosphere back to being O-rich.

AGB stars are also characterized as slow pulsators and over time these pulsations lift material above the photosphere which condenses into dust. Similarly, red supergiants (RSGs) are massive stars that produce dust shells as they near the end of their lives. In the most intense cases, some stars can produce enough dust to become hidden in the optical and near-Infrared (NIR) due to self-obscuration. In the case of AGB stars, they are designated as ``extreme'' AGB (XAGB) stars. These rare stars contribute a significant fraction of the dust found in a galaxy (e.g., Riebel et al. 2012; Boyer et al. 2012).

In order to see behind the veil these stars often establish, it becomes necessary to begin to observe at longer wavelengths than the optical and NIR. Observations in the mid-IR and far-IR enable the behavior of both the central stars and their shells to be better characterized. However, extragalactic studies at these wavelengths have been lacking because the combination of large distances and beam sizes on early ground and space-based IR telescopes made distinguishing even the most luminous point sources difficult.  Studies of mid-IR variability in extragalactic sources had been predominantly restricted to the Magellanic Clouds (van Loon et al. 1998, and references therein) due to their relative proximity to the Galaxy. 

The {\it Spitzer Space Telescope} (Werner et al. 2004) observed across the IR spectrum from 3.6 to 160~\micron\ and was able to provide better angular resolution and sensitivity than any prior ground or space-based facilities. The Infrared Array Camera (IRAC, Fazio et al. 2004), Multiband Imaging Photometer for {\it Spitzer} (MIPS, Rieke et al. 2004), and the Infrared Spectrograph (IRS, Houck et al. 2004) on {\it Spitzer} improved the ability to resolve embedded point sources compared to previous instruments due to smaller beam sizes and better sensitivity. Further, the {\it Herschel} Space Observatory ({\it Herschel}, Pilbratt et al. 2010) has allowed for improved space-based resolution of extragalactic targets in the far-IR and sub-mm to detect the most obscured objects.

{\it Spitzer} was able to study these enshrouded AGB and RSG stars, revealing evidence that variability in the mid-IR and far-IR is more common than previously expected (McQuinn et al. 2007; Mould et al. 2008; Vijh et al. 2009; Riebel et al. 2010; Sloan et al. 2010; Boyer et al. 2010). A study of IR variability in the LMC was done using two epochs of IRAC and MIPS 24~\micron\ photometry separated by 3 months (Vijh et al. 2009). About 2,000 variable sources were found from this relatively short baseline among the millions of point sources in the SAGE catalog (Meixner et al. 2006). These variable sources, listed in Table 3 of Vijh et al., are mostly AGB star candidates of all classes (O-rich, C-rich, and extreme), along with a few massive young stellar object (YSO) candidates. These findings have allowed a better understanding of the enrichment of the interstellar medium (ISM) for a diverse set of environments and metallicities.

M33 is a late-type spiral galaxy and the third largest member of the Local Group. It was also the target of multiple {\it Spitzer} observations over the length of the cold mission with IRAC and MIPS. McQuinn et al (2007; hereafter McQ07) conducted the only previous study of mid-IR variability in M33 with five epochs of IRAC observations. This study found 2,923 variable stars from a point source catalog of 37,650 objects. These variables are predominantly AGB stars, both carbon (C-rich) and oxygen (O-rich) types, and $80\%$ of the AGB stars detected at 8~\micron\ in M33 are surrounded by dust shells. A total of 515 discrete MIPS 24~\micron\ sources have been identified (Verley et al. 2007). Verley et al. applied a series of theoretical and observational diagnostics and concluded that the population of sources was comprised mainly from supernova remnants (SNRs) and HII-regions.

In this work, we report the first use of variability at 24~\micron\ to identify evolved stars in M33. In \S2, we outline the multi-wavelength observations that we used to help determine the source of the variability. In \S3, we detail the method for source extraction, how our variable candidates (VCs) were determined, and their association with M33. In \S 4, we discuss our efforts to identify our targets at other wavelengths and the SED modeling that we applied. In \S 5, we highlight sources that we were able to classify, and present the results of our SED modeling. The results are discussed in \S6. Finally in \S7, we summarize the main conclusions of this work.

\section{Observations \& Reduction}
\subsection{MIPS 24~\micron}
The raw MIPS 24~\micron\ data (PID:5, PI: Gehrz) were retrieved from the {\it Spitzer} Heritage Archive. There were four epochs of M33 observations, which were spaced irregularly in time over ~740 days. The epochs of the observations are 2003 December 29-30, 2005 February 3, 2005 September 5, and 2006 January 9-12. The individual epoch data were processed using the MIPS DAT package (Gordon et al. 2005), and calibrated as laid out by Engelbracht et al. (2007).

\subsection{Additional {\it Spitzer} Observations}
As stated in the introduction, the most extensive survey for variables at IR wavelengths in M33 was conducted by McQ07. We cross matched our VCs and the McQ07 variable and non-variable catalogs. A tolerance of 3\arcsec\ was used to find any matches, to reduce the potential number of sources that can be blended together at 24~\micron. Twenty-three of our VCs were found in a total of 25 matches with 19 in the variable catalog and 6 in the non-variable catalog. A counterpart for VC 1 was not found in McQ07. Two of our sources, VC 4 and VC 24, had matches in both catalogs, but since both objects fall within the 24~\micron\ beam size we are inclined to believe that the variable object is the counterpart to our VC. The three bands of IRAC photometry (3.6, 4.5 and 8.0 $\mu$m) published by McQ07 were incorporated into our SED modeling.

Archival four-band IRAC and MIPS 70~\micron\ images for M33 were also retrieved from the Local Volume Legacy Survey (LVL, Dale et al. 2009). StarFinder (Diolaiti et al. 2000) catalogs were generated for these observations. An archival IRS spectrum was found and retrieved for only one object, VC 7 (AOR 16212736, PI Gehrz), from the Cornell Atlas of Spitzer IRS Sources (CASSIS\footnote{CASSIS is a product of the Infrared Science Center at Cornell University, supported by NASA and JPL.}; Lebouteiller et al. 2011). VCs 1, 2, 3, 4, and 11 were matched to possible counterpart point sources in the MIPS 70~\micron\ observations. We were able to recover 19 of our VCs in all four IRAC bands. The LVL data are combined multi-epoch observations of the McQ07 data, which allows for fainter/more obscured sources to be detected, and the inclusion of the IRAC 5.8~\micron\ band, which McQ07 did not include in their analysis. In addition, 15 of our VCs are also found in the 24~\micron\ observations of Verley et al. (2007).

\subsection{{\it Herschel} Observations}
M33 observations were taken as a part of the {\it Herschel} Open Time Key Program HERM33ES ({\it HERschel} M33 Extended Survey; Kramer et al. 2010). The mapping was done in parallel mode with both the Spectral and Photometric Imaging REceiver (SPIRE) at 250, 350, and 500~\micron\ (Griffin et al. 2010) and the Photodetector Array Camera and Spectrometer (PACS) at 70, 100, and 160~\micron\ (Poglitsch et al. 2010). The SPIRE data for M33 were retrieved from the {\it Herschel} Science Archive after it became publicly available, and were processed through the KINGFISH pipeline (Kennicutt et al. 2011). M33 PACS maps, at 100 and 160~\micron, were kindly provided by the HERM33ES team (Kramer, private communication). StarFinder catalogs were generated for the {\it Herschel} observations and our VCs were cross matched against these point sources. VCs 1, 2, 3, 4, and 11 had matches in both PACS 160~\micron\ and  SPIRE 250~\micron\ observations. The PACS 100~\micron\ image is not as sensitive as the 160~\micron\ image, which resulted in only 2 candidates being recovered. The longer wavelength SPIRE 350~\micron\ and 500~\micron\ observations only found 2 and 1 matches, respectively, as the increasing beam sizes quickly led to a reduction in the resolution of the observations. PACS 70~\micron\ mapping (PI: M. Boquien) of M33 was not as deep as the HERM33ES mapping. 

\subsection{UV, Optical, \& IR Data}
The WISE All-Sky Data Release (Cutri et al. 2012) was queried for our sources as it has similar wavelength coverage. We were able to recover 23 out of 24 of our VCs in all 4 WISE bands. GALEX (Galaxy Evolution Explorer, Martin et al. 2005) observations were obtained from Gil de Paz et al. (2007). M33 ground-based optical images in B, H$\alpha$, and R were retrieved from the Local Group Survey (LGS, Massey et al. 2006). Combined mosaics from the Two Micron All-sky Survey (2MASS, Skrutskie et al. 2006) were obtained through the Large Galaxy Atlas (Jarrett et al. 2003). A complete summary of all available data can be found Table 1. The table contains telescope, instrument and wavelength/filter, and reference paper for that particular observation.

We also searched for counterparts in two ground-based M33 variable surveys using the VizieR Catalogue Service provided by the Strasbourg Astronomical Observatory. The first, completed by Hartman et al. (2006), observed M33 with MegaPrime/MegaCam in the g$'$, r$'$, and i$'$ bands using the 3.6-m Canada-France-Hawaii Telescope (CFHT). They identified over 36,000 variables that are overwhelmingly comprised of evolved stars. Their observations are broken up into two categories: point sources (Tables 2-4) and extended sources (Tables 5 \& 6). A search for counterparts in the former found VCs 7, 10, 14 (see \S 4.4.1), 17, and 24 have at least one known variable within 3\arcsec, with VC 7 having two. Matches to VCs 3 (1 match), 4 (2 matches), and 24 (5 matches) were found in the latter table with the same tolerance. The cases of multiple matches, especially VC 24, highlights an issue with the area covered by the MIPS 24~\micron\ beam. Hartman et al. published semi-instrumental magnitudes, therefore this photometry is not included in our SED modeling in \S4.3. The second survey searched consisted of $JHK_S$ observations conducted with Wide Field Camera (WFCAM) on 3.8-m United Kingdom InfrarRed Telescope (UKIRT), which were completed by Cioni et al. (2008). They follow-up on variables monitored by Hartman et al. (2006) and by McQ07 in order to obtain their NIR colors for AGB classification. VCs 13, 17, 18, and 24 were recovered by Cioni et al.

Our VCs were also cross matched to the catalog of a NIR monitoring survey for variable stars across the entire disk of M33 by Javadi et al. (2014). VCs 6, 8, 9, 13, 16, 21, and 23 were matched to strongly confident variable sources with J--K$_s >$ 2.4 mag. Additionally, VCs 10, 17, and 20 were matched to sources likely to be variable, but with less confidence due to fewer epochs. These also have J--K$_s >$ 2.4 mag. The remaining VCs, excluding 1, 5, 7 and 14, which are discussed in  their respective sections, are not found to be variable by Javadi et al. This implies that these VCs are possibly not evolved stars (see discussion in \S6).

\section{Variable Identification}
Catalogs of point sources from individual epochs were extracted with both PSF photometry using the IDL routine StarFinder, and aperture photometry using SExtractor (Bertin \& Arnouts 1996). The aperture size was selected to capture $90\%$ of the flux. While holding Epoch 1 as the reference, sources separated by a maximum of 3\arcsec\ (one-half of the 24~\micron\ beam FWHM) between epochs were assumed to match. The search algorithm included an adaptive tolerance in order that the nearest neighbor was always selected. The final matched StarFinder generated catalogs resulted in 2,645 sources from all epochs, while the matched SExtractor catalogs contained 1,333 sources. The average separation for our sources between epochs was less than 1.0\arcsec.

A reduced $\chi^2$ statistic was then calculated for the sources in each matched catalog to help determine any signatures of variability. The statistic is defined as 
\begin{displaymath}
\chi^2_{24~\micron} \equiv 
\frac{1}{N_{obs}} 
\times 
\left [ \sum^{N_{obs}}_{j=1} \frac{({F_{24~\micron}}_j - \langle F_{24~\micron} \rangle)^2}{{\sigma^2_{24~\micron}}_j} \right ],
\end{displaymath}
where ${F_{24~\micron}}_j$ is the 24~\micron\ flux and ${\sigma^2_{24~\micron}}_j$ the associated uncertainty for the {\it j}th observation, $\langle F_{24~\micron} \rangle$ the average 24~\micron\ flux, and $N_{obs}$, the total number of observations. A Gaussian was fitted to the distribution of $\chi^2$ with any values $\ge 3\sigma$ taken as evidence of variability. We found a total of 428 from the matched StarFinder catalogs and 155 candidates from the SExtractor catalogs.

The Gaussian fits of the $\chi^2$ distribution were examined for any overestimation or underestimation of the uncertainties, which would determine if the calculated $\chi^2$ statistic is a good indicator of variability. The distribution of the StarFinder extracted photometry peaked at a few tenths, which suggests that we have overestimated the associated uncertainties. The distribution for the SExtractor photometry peaks around 2, which suggest an underestimation of the uncertainties. A final matching between these catalogs was done to make a list of strongly suspected variable sources. To ensure we are matching identical sources in the two catalogs, we required a stricter tolerance of 1\arcsec. This resulted in 24 variable candidates (VCs) which are listed in Table 2. This extreme culling is most likely the product of the strength and weakness of the two methods of photometry that we used. PSF photometry is going to do much better than aperture photometry in both crowded and high background regions of the images, which is reflected in the total number of sources from each method. While aperture photometry provides a more stable centroid to limit  false variability that can occur with the PSF method, which is demonstrated by the lower percentage of source identified as variable by aperture ($\sim$12\%) versus PSF ($\sim$16\%). Therefore we choose to report only our PSF photometry, since the aperture photometry merely served as an additional variability check.

The locations of the VCs are shown in Figure 1. The Holmberg radius (8.7 kpc) for M33 is plotted on the figure to help determine the likelihood that the identified targets are associated with the galaxy. All 24 candidates are found to be located within the Holmberg radius. The proposed variables are described in Tables 2 and 3. The VC number is in column 1, the right ascension and declination (J2000) are given in columns 2 and 3, the average 24~\micron\ flux and RMS error are in column 4, the PSF photometry amplitude is given in column 5, and finally the PSF and aperture reduced $\chi^2$ statistics are found in columns 6 and 7, respectively. Individual stellar photometry is contained in Table 3. The 24~\micron\ light curves for our targets can be found in Figure 2. 

\section{GRAMS SED Modeling}
In order to gain a better understanding into the nature of our VCs, we placed them in the context of the LMC's variable population, which is much better studied than in M33. To this aim, we brought our VCs to the distance of the LMC. This was done assuming an M33 distance of 840 kpc (Freedman et al. 1991), and an LMC distance of 50 kpc (e.g. Feast 1999). Figure 3 shows our sample plotted in a [8.0] vs [8.0]--[24.0] CMD with the LMC variables (Vijh et al. 2009). The lines drawn in Figure 3 mark the region where YSOs have the highest probability of being found (Whitney et al. 2008). Our VCs fall outside of this boundary indicating that they are most likely dusty, evolved stars.

Given the results of our Spitzer CMD and cross-matching above, we fit the VCs SEDs with the Grid of RSG and AGB ModelS (GRAMS). The GRAMS grid consists of radiative transfer models for oxygen-rich (silicate) and carbon-rich (amorphous carbon and silicon carbide) dust around AGB/RSG stars (Sargent et al. 2011; Srinivasan et al. 2011). The best-fit GRAMS models are able to separate evolved stars into O-rich and C-rich types (Riebel et al. 2012; hereafter R12).

We computed best-fit models of O-rich and C-rich types by performing chi-square fits to the observed SEDs. The fits are shown in Figure 9. For each source, we fit the optical through MIPS 70~\micron\ data. First, we incorporated the variability information from McQ07 into the IRAC and WISE flux uncertainties, and the variability from our multi-epoch observations into the MIPS24 uncertainties. In addition, we inflated the uncertainties of the optical and near-infrared fluxes in a manner similar to R12. As AGB variability has a stronger effect on the flux at shorter wavelengths, it is important to account for this variation by decreasing the weight given to this wavelength range when computing the chi-square. Further, the WISE W3 band's location between the IRAC and MIPS bands allows it to probe the strength of the silicate feature in O--rich AGB stars or RSGs, as well as the silicon carbide feature seen in carbon stars. As a consequence, this WISE band can better constrain the optical depth of the best-fit GRAMS model.

We accounted for variability in each band by adding a term in quadrature to the photometric uncertainty. For the WISE bands, this term was determined in such a way that the relative uncertainty was equal to that of the multi-epoch observation in the nearest IRAC band for that source. If a source lacked multi-epoch data, we computed this term using the median relative uncertainty due to variability in that band. Six of our sources had matching near-infrared photometry. In these bands, we used a relative uncertainty equal to twice that of the IRAC 3.6 $\mu$m band, because the amplitude of variability is larger at shorter wavelengths.

\section{Individual Variable Sources}
\subsection{VC 1: NGC 604}
The location of VC 1 happens to correspond with the position of the giant HII region NGC 604 ($\alpha$: 01:34:33.56, $\delta$: +30:47:03.5) in M33. Therefore it is highly unlikely the fluctuations of a single source are being followed given the combination of the beam size covering a large area, and the complexity of NGC 604 (Eldridge \& Rela{\~ n}o 2011 and references therein). This explains why the parameters for the GRAMS modeling (see below) of VC 1 are at or near their maximum limits, and a large IR color excess was found, see Figure 9 and Table 4. Javadi et al. (2014) detect several NIR variables within NGC 604. However, given the difficulties of this region the true source of the variability at 24~\micron\ remains unknown at this time.

\subsection{VC 5: EAGB-2}
A search through archival IRAC observations for nearby galaxies was performed by Khan et al. (2010). The goal of Khan et al. was to find as many self-obscured massive stars in these galaxies in order to determine progenitors for SN2008S-like transients. IRAC photometry of XAGB (EAGB in Khan et al.) stars in M33 is published in Table 1 (Khan et al. 2010). The position of VC 5 matches their second entry, referred to as ``EAGB-2'', to 0.7\arcsec. EAGB-2 was deemed a ``Class-A'' object by Khan et al., because it was detected in both IRAC 3.6 and 4.5~\micron\ imaging. Our {\it Spitzer} color-color (CC) diagram analysis, Figure 10, indicates that VC 5 should be classified as O-rich. This is supported by the best-fit GRAMS modeling, see Figure 9 and Table 4. Previous simple blackbody SED fitting estimated the bolometric luminosity of $\sim 10^4$ L$_\odot$ (Khan et al. 2010). Our best-fit GRAMS modeling for the same parameter is $2.51 \times 10^4$ L$_\odot$, where the lack of adequate error bars is due the coverage in parameter space (see \S 6). Figure 4 contains 8 postage stamp images covering a 1\arcmin\  $\times$ 1\arcmin\ region surrounding VC 5 from NUV to the sub-mm. The best-fit GRAMS dust production rate (DPR) is $(7.48 \pm 1.52) \times 10^{-7}$ M$_\odot$ yr$^{-1}$ and $\tau_{1\micron}$ on the order of 30, which are among the highest for our VCs. Further, Khan et al.'s as well as our own color-based extreme AGB classification demonstrates the problem in assuming that AGB stars with very red colors are most likely C-rich (see \S 6). This object was found by Javadi et al. (2014) to not be variable in their NIR monitoring of M33. 

\subsection{VC 7: SSTM3307 J013412.95+302938.1}
The location of VC 7 is consistent with the identified IRAC variable SSTM3307 J013412.95\\+302938.1 (McQ07). Postage stamp images covering a 1\arcmin\  $\times$ 1\arcmin\ region surrounding VC 7 from NUV to the sub-mm are shown in Figure 5. It has been further classified by Polomski et al. (in prep) as an ``extreme'' Mira, and M33 analog of OH26.5+.6 (Garc{\'{\i}}a-Hern{\'a}ndez et al. 2007, Sylvester et al. 1999) through an analysis of an IRS spectrum. The spectrum shows a deep silicate absorption feature at 10 $\mu$m, which allows us to better constrain the parameters of our GRAMS best-fit model, see Figure 9. The best-fit parameters are a luminosity of $3.37 \times 10^4$ L$_\odot$ and DPR of $(6.13\pm1.22) \times 10^{-7}$ M$_\odot$ yr$^{-1}$. The high obscuration suggested by the value of $\tau_{1\micron}$ for this object (see Table 4) suggest that those objects are our detections at IRAC and MIPS. VC 7 was found to be O-rich by the GRAMS modeling. This reinforces the conclusions presented by Polomski et al. (in prep) that this an extreme, enshrouded O-rich Mira-type variable in M33.

The star that is closest to the 24~\micron\ and IRAC centroid is HBS 220970 (Hartman et al. 2006), which has detections in both the r$'$ and i$'$ bands. Javadi et al. (2014) find a non-variable star with moderate red colors, which they believe to be the NIR counterpart to HBS 220970. The i$'$ light curve from the Hartman et al. survey is presented in the left side of Figure 6 and a folded light curve, assuming the Polomski et al. classification, on a period of 142.857$\pm$0.7568 days on the right. The period was determined by fitting a sine wave function of the form 
\begin{displaymath}
v(\phi) = \gamma + K_{2}sin(2\pi\phi + \psi),
\end{displaymath}
where $\phi$ is the phase, $\psi$ is the phase shift, $K_2$ the amplitude, and $\gamma$ is the systemic shift. In the first iteration, all of the parameters were allowed to vary and then subsequent iterations set by the minimum $\chi^2$ of this iteration.

While our determined period might be on the low end for a dust enshrouded AGB star, the pulsations seen in the IR, which are following the circumstellar material, have to be following the behavior of the central star. Thus, the {\it Spitzer} observations were then folded on this period and can be seen in Figure 7. The MIPS and IRAC observations were started during the beginning of a gap in the Hartman et al. (2006) monitoring at JD 2453000, and continued until past the ground-based campaign completed at JD 2453771.16. The four MIPS epochs (red squares) and six IRAC 8.0~\micron\ epochs (green squares) were plotted on this fold in order to see if there was any delay, which would correspond to a delay between the intrinsic stellar pulsation and response by the dust shell. No such delay was found. However, an unexpected result comes from the sixth epoch of IRAC observations, which is significantly fainter than expected for a purely pulsation decline. 

The final IRAC epoch was not included in the variability determination by McQ07, since the observations were taken in the advanced stages of their analysis. The post-Basic Calibrated Data (PBCD) of the sixth IRAC epoch containing VC 7 (AOR 16045568, PI Gehrz) were retrieved from the Spitzer Heritage Archive, and a StarFinder catalog was generated for 8.0~\micron\ mosaic. Only the 8.0~\micron\ band was selected in order to correspond to the Polomski et al. analysis of the IRS data, which determined the 8.0~\micron\ magnitude from the SL2 component of the IRS spectrum. The Polomski et al. (in prep) calculated magnitude and our StarFinder magnitude are in agreement, and expected since the IRS observation was made at JD 2453755.67 while the sixth IRAC epoch was taken roughly 15 days later at JD 2453771.16. Both the additional IRAC epoch and IRS spectrum integration appear to point to a non-pulsational decline occurring during the time of these observations.

\subsection{VC 14: VHK 71}
The position of this source coincides with the variable star VHK 71 (van den Bergh et al. 1975). The General Catalogue of Variable Stars (Samus et al. 2013) gives a variable type of ``SRC'', which defines semi-regular late type supergiants of the $\mu$ Cephei type with periods of 30 to several thousand days. This classification was determined by spectroscopic observations by Giovagnoli \& Mould (1994) that found VHK 71 to have an M2 spectral type. A variable source coinciding with the position of VHK71 is found by both McQ07 and our work, while Javadi et al. (2014) do not find it to be variable in ground-based NIR photometry. Our CC diagram analysis (Figures 12 and 13) and best-fit GRAMS modeling (Figure 9) both find that the best chemical classification for VC14 would be an O-rich atmosphere. It is important to recall that RSGs are folded into the GRAMS O-rich designation. Further examination of the best-fit GRAMS parameters suggests that the bolometric luminosity is $(2.09 \pm 0.29) \times 10^5$ L$_\odot$ with an effective temperature of 3700 $\pm$ 600 K. These values translated onto an H--R diagram would categorize VC 14 as an RSG star. They are also in close agreement with values for the same parameters, L $=$ $1.41 \times 10^5$ L$_\odot$ and T$_{\rm eff} \sim$3400 K, recently determined in the literature (Drout et al. 2012). 

The period of VHK 71 is 760 days (Kinman et al. 1987), and the MIPS coverage for VC 14 spans $\sim$750 days. The 24~\micron\ light curve, which is shown with the other VCs in Figure 2, does appear to closely follow the period determined by Kinman et al. (1987), if the first epoch is taken as the reference point. VC 14/VHK 71 was also detected and cross-matched in other ground-based surveys of M33 such as Macri et al (2001), Hartman et al. (2006), and the LGGS (Massey et al. 2006, 2007), and Javadi et al. (2014). The detections at shorter wavelengths (see Figure 8 for postage stamp images) indicate the presence of an optically thin circumstellar envelope, which is reinforced with the best-fit GRAMS DPR, $(1.39 \pm 0.74) \times 10^{-8}$ M$_\odot$ yr$^{-1}$, low value for $\tau_{1\micron}$, and the NIR photometry of Javadi et al. (2014). 

\section{Discussion} 
GRAMS fitting allows for constraints to be placed on the following central star and circumstellar shell properties: elemental enrichment (C-rich vs O-rich), luminosity, dust-production rate (DPR), and the optical depth of the shell at 1~\micron\ ($\tau_{1\micron}$). The best-fit output parameters for our VCs can be found in Table 4 and on their SEDs in Figure 9. The uncertainties on our GRAMS parameters were determined from the 250 best-fit models for the given GRAMS classification by computing the median absolute deviation from the median (MADM). However, there are some cases that result in an undetermined uncertainty for some parameters because all 250 models for those sources collapse to the same value. This is represented as ``0.00'' in Figure 9 or a lack of quoted uncertainty elsewhere. We can place these values, ignoring VC 1 (NGC 604, see \S5.1) in the context of AGB and RSG stars in the much better-studied LMC.

Our GRAMS modeling found that, of the remaining 23 VCs (2 to 24), 20 sources can be classified O-rich, 1 as C-rich, and that the classification for VCs 6 and 22 is unclear (O and C chi-squares are within 5\% of each other; see Figure 10). We adopt the O-rich classification and parameters for the two unclear sources, since in both cases the lower $\chi^2$ is the O-rich model. RSGs are also folded into the O-rich designation. These findings are a radical departure from what we would have found if we had used stellar properties presented in Table 5 of McQ07. The presence of a dust shell is indicated by McQ07 if a star has [3.6] -- [8.0] $>$ 0.5, which all of our VCs have. If we were to follow their distinction for O-rich ([3.6] -- [4.5] $<$ 0.2, [3.6] -- [8.0] $<$ 0.5), and C-rich ([3.6] -- [4.5] $<$ 0.2, [3.6] -- [8.0] $<$ 0.5) AGBs, we would have to conclude that the overwhelming majority of our VCs are C-rich. However, identification based on IRAC colors alone are subject to a strong degeneracy between carbon stars and XAGB stars, which are a mixture of very enshrouded C-rich {\it and} O-rich AGB stars. This overlap is illustrated in Figure 11, where the majority of our VCs (pink diamonds) fall in a regime where both O-rich (blue dots) and C-rich (red dots) overlap. Statistically almost all extreme AGB stars are C-rich (R12), which is most likely why McQ07 made their selection criteria as such. Our 24~\micron\ photometry can be used to break this degeneracy, as shown in Figure 12. However, it is important to note that our GRAMS modeling are not sufficient enough to overcome the degeneracies between O-rich stars and other types of objects (e.g., YSOs).

The average bolometric luminosity of the VCs is $3.8 \pm 0.9 \times 10^4$ L$_\odot$, with a minimum value of $(5.19 \pm 0.38) \times 10^3$ L$_\odot$ and maximum value of $(2.09 \pm 0.29) \times 10^5$ L$_\odot$. The range of luminosities estimated by R12 from their GRAMS SED fitting for LMC AGB candidates was $\sim$ 2 $\times$ 10$^3$ -- 2.5 $\times$ 10$^4$ L$_\odot$ for carbon stars and $\sim$ 10$^3$ -- 2.5 $\times $10$^5$ L$_\odot$ for O-rich AGB stars. However, the small number of sources ($<$30) with luminosities above the classical AGB limit (M$_{\rm bol} <$ --7.1 mag, L $>$ 54,000 L$_\odot$) are probably supergiants (see Figure 14 and \S 4.2 in R12). This suggests that VCs 2, 3, 4, 14, and 17 are RSGs. VC17, in particular, is found by Javadi et al. (2014) to be consistent with a dusty RSG star, as indicated by our GRAMS SED modeling. However, VC 4 is only 5\% more luminous than this upper limit. It is possible for massive O-rich AGBs undergoing HBB to overcome the classical AGB limit (e.g., Bl{\" o}cker \& Sch{\" o}nberger 1993); the brightest OH/IR stars are examples of this.

The total DPR for our VCs is $(2.6 \pm 0.2) \times 10^{-5}$ M$_\odot$ yr$^{-1}$. This number is comparable to the global DPR of $(2.13 \pm 0.02) \times 10^{-5}$ M$_\odot$ yr$^{-1}$ derived by R12 for the {\em entire AGB/RSG population} in the LMC. In order to determine if our VCs fit in with these previous results we can place the VCs onto Figure 16 from R12, which examined DPR vs bolometric luminosity by chemical type. In Figure 13, VCs are shown as squares and background points are from R12. with blue and red representing O-rich and C-rich, respectively. Our sources are on average both brighter and produce more dust than the LMC XAGB stars studied by R12. We also find that the most extreme dust producers are in fact O-rich and not C-rich. Massive O-rich AGB stars can attain DPRs of up to a few times 10$^{-7}$ M$_{\odot}$ yr$^{-1}$ during the superwind phase; the total return rate from our VCs is therefore not too surprising.

Dust budget estimates in the Magellanic Clouds point to a deficit (e.g., Matsuura et al. 2009; Boyer et al. 2012; Matsuura et al. 2013) compared to the dust mass observed in the interstellar medium (ISM). R12 found that $\sim$ 75\% of their DPR came from their XAGB candidates, which comprised only $\sim$ 4\% of the total evolved star population of the LMC. It is possible that heretofore undetected sources such as our VCs contribute a non-negligible amount of dust to the ISM, as is demonstrated by comparing our total DPR to the value of Javadi et al. (2013), 2.25 $\times$ 10$^{-5}$ M$_\odot$ yr$^{-1}$, for the inner square kpc of M33. The significant difference in our studies is the use of any data longer than 8~\micron\ to find the DPR. Javadi et al. did not include wavelengths beyond 8 $\mu$m due to significant crowding in the central square kpc in M33. It becomes very difficult to accurately constrain the DPR for extremely dusty sources, such as ours (see Figure 14), without information from both sides of the 10~\micron\ silicate feature. Thus, the inclusion of MIPS 24~\micron\ photometry in our determination for the DPR of our sources, despite the large physical area that the beam covers at the distance of M33, enables us to at least use our findings as an upper bound -- even if we are likely overestimating the true value. 

Our GRAMS modeling results are consistent with the above, showing that our sources are among the brightest, and by extension very likely, the most massive, evolved stars in M33. Stars in this stage of their evolution are known to be regular to irregular pulsators, which drives their brightness fluctuations. It is also well known that there is an inverse relationship between wavelength and the amplitude of the observed fluctuations. Therefore, at 24~\micron\ we are seeing these intrinsic changes reprocessed through the warm circumstellar material. This opens the possibility that future high resolution space-based IR missions, like JWST, will be able to reveal additional objects like our VCs in nearby galaxies. 
 
\section{Conclusion}
We have conducted a search for variable sources in M33 with archival MIPS 24~\micron\ observations. Our conservative analysis has uncovered 24 variable candidates (VCs) from thousands of objects from catalogs generated both with PSF and aperture photometry. These VCs are the first known instances of 24~\micron\ variability detected in any galaxy beyond the Magellanic Clouds (MCs). Using prior studies in the LMC as an example, we acquired a suite of archival observations, ranging from the far UV to the sub-mm, with which we performed: visual inspection, GRAMS SED modeling, and analysis of {\it Spitzer} color--magnitude and color--color (CC) diagrams of all of our VCs. Aside from VC 1, coincident with the location of NGC 604, we have determined that our VCs are very likely dusty, massive evolved stars. Our GRAMS modeling suggests that 20 are oxygen-rich (O-rich), 1 carbon-rich (C-rich), and 2 have uncertain classification. The O-rich classification can be further refined by a cut at the classical AGB luminosity limit resulting in at least 4 RSGs. The remaining sources can be classified as ``extreme'' AGB stars, which are characterized by significant obscuration. We were able to further cross match 3 VCs to known M33 sources in the literature. These are: VC 5 -- an XAGB star (Khan et al. 2010); VC 7 -- an extreme Mira (Polomski et al. in prep); and VC 14 -- VHK 71 (van den Bergh et al. 1975). The average bolometric luminosity is $(3.8 \pm 0.9) \times 10^4$ L$_\odot$, reinforcing that these objects are evolved stars, while the total DPR value for our VCs was found to be $(2.3 \pm 0.1) \times 10^{-5}$ M$_\odot$ yr$^{-1}$. This is certainly an underestimate of the true value in M33, but our VCs are most likely a large contributor to this value. We fully expect that the JWST-era will enable much better resolution of stars in these late stages of evolution to even further extragalactic distances.

We would like to thank our referee, Jacco van Loon, for his valuable comments for sharing the results from his group's upcoming variability paper, which helped improve our discussion. This work is based on observations made with the Spitzer Space Telescope, which is operated by the Jet Propulsion Laboratory, California Institute of Technology under a contract with NASA. Based on observations made with the NASA/ESA Hubble Space Telescope, and obtained from the Hubble Legacy Archive, which is a collaboration between the Space Telescope Science Institute (STScI/NASA), the Space Telescope European Coordinating Facility (ST-ECF/ESA) and the Canadian Astronomy Data Centre (CADC/NRC/CSA). This publication makes use of data products from the Two Micron All Sky Survey, which is a joint project of the University of Massachusetts and the Infrared Processing and Analysis Center/California Institute of Technology, funded by the National Aeronautics and Space Administration and the National Science Foundation. This publication makes use of data products from the Wide-field Infrared Survey Explorer, which is a joint project of the University of California, Los Angeles, and the Jet Propulsion Laboratory/California Institute of Technology, funded by the National Aeronautics and Space Administration.

\begin{figure}
\begin{center}
\epsscale{1.0}
\plotone{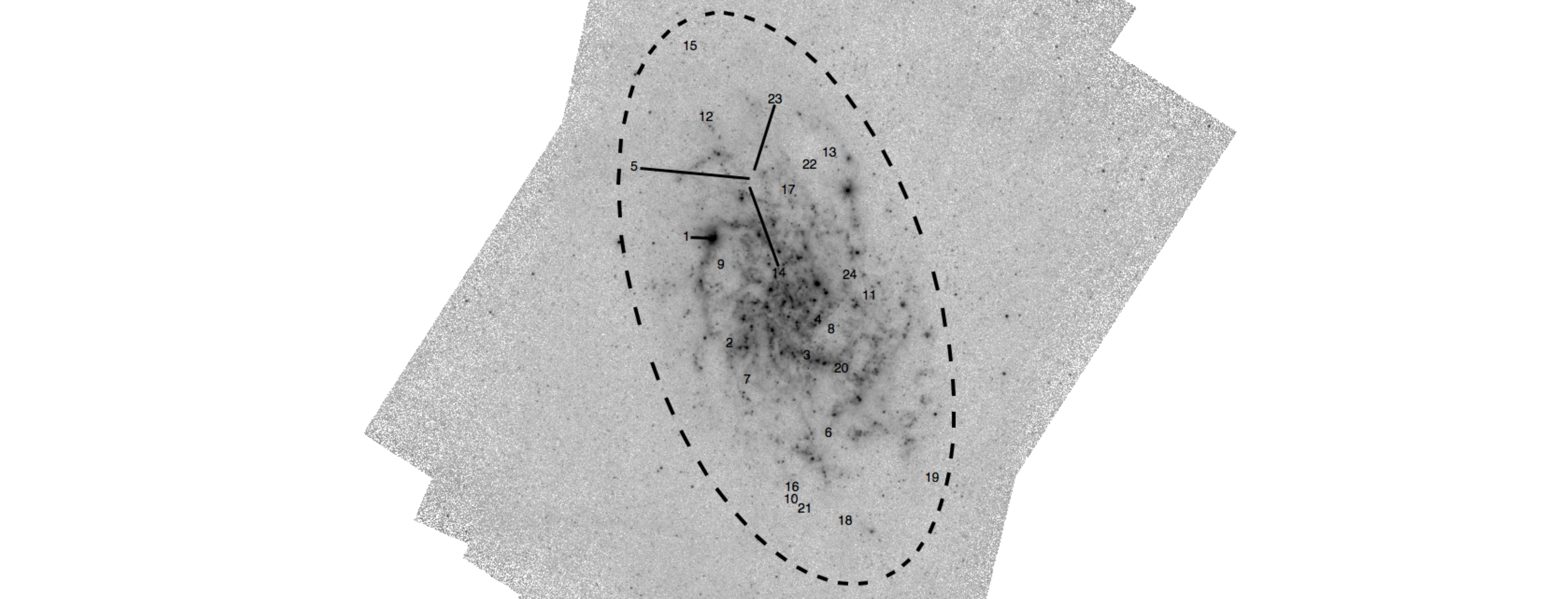}
\caption{The MIPS 24~\micron\ image for M33. The field of view is roughly $1.4\arcdeg \times 1.2\arcdeg$. The dashed ellipse represents the Holmberg radius M33 (8.7 kpc). The numbers represent the locations of the variable candidates (VCs) and correspond to their positions given in Table 2.}\label{fig1}
\end{center}
\end{figure}

\clearpage

\begin{figure}
\begin{center}
\epsscale{1.0}
\plotone{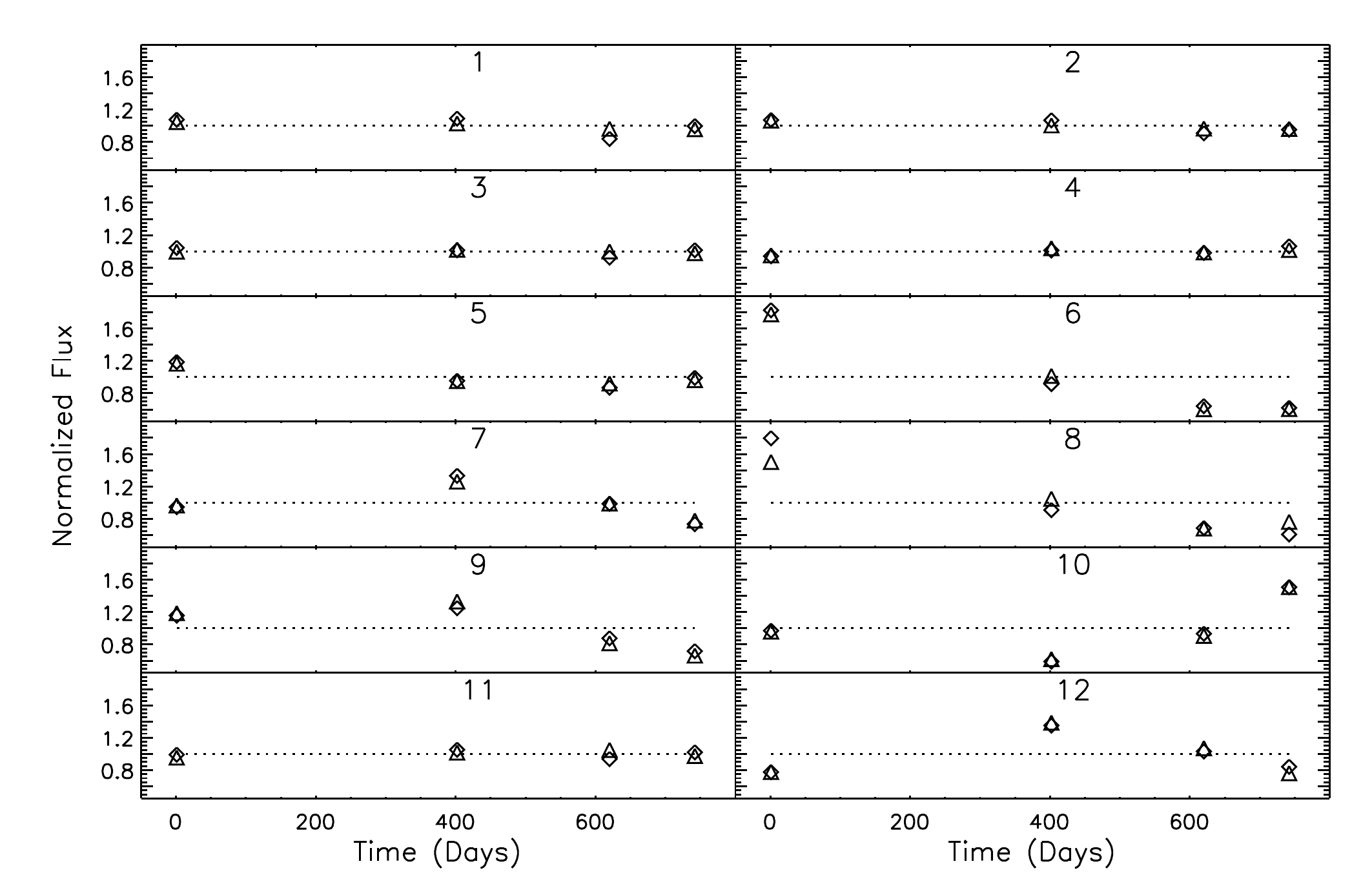}
\vspace*{-10truemm}
\caption{The 24~\micron\ light curves for VCs 1-12. Diamond points are normalized PSF photometry and triangles are normalized aperture photometry. A dotted line is drawn through the average flux of the measurements. The error bars are smaller than the points.}\label{fig2}
\end{center}
\end{figure}

\setcounter{figure}{1}

\begin{figure}
\begin{center}
\epsscale{1.0}
\plotone{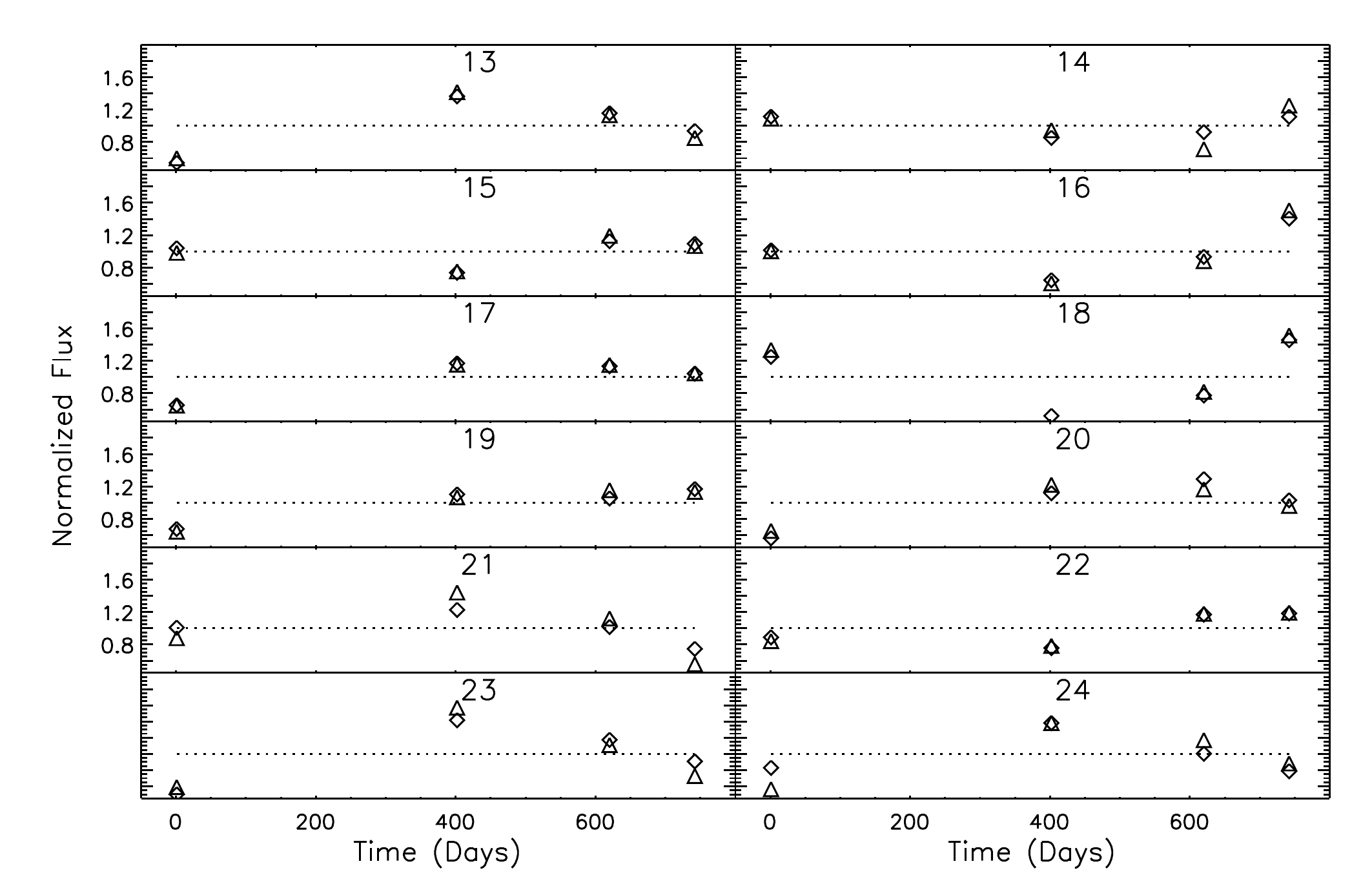}
\vspace*{-10truemm}
\caption{Figure 2 continued: the 24~\micron\ light curves for VCs 13-24.}\label{fig2}
\end{center}
\end{figure}

\clearpage

\begin{figure}[t]
\epsscale{1.0}
\plotone{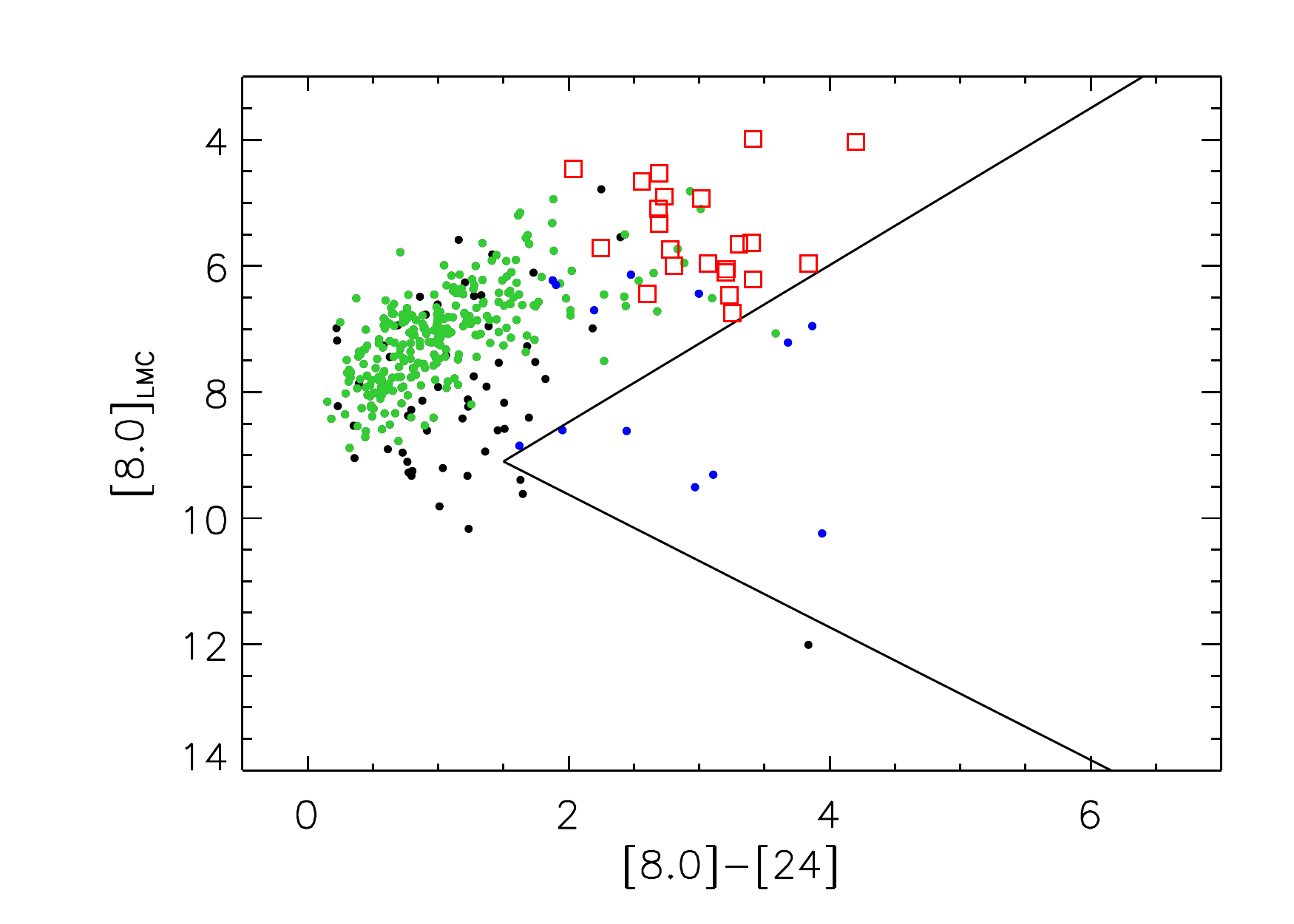}
\vspace*{-5truemm}
\caption{CMD showing [8.0] vs [8.0]--[24], where the green points are 24~\micron\ variable XAGB candidate stars, blue points are 24~\micron\ variable massive YSO candidates, and black dots are other sources from Vijh et al. (2009). Our VCs are represented by the red squares. The lines that have been drawn indicate the region to the right of which YSOs are most likely to be found (Whitney et al. 2008).}\label{fig3}
\end{figure}

\clearpage

\begin{figure}[t]
\epsscale{1.0}
\plotone{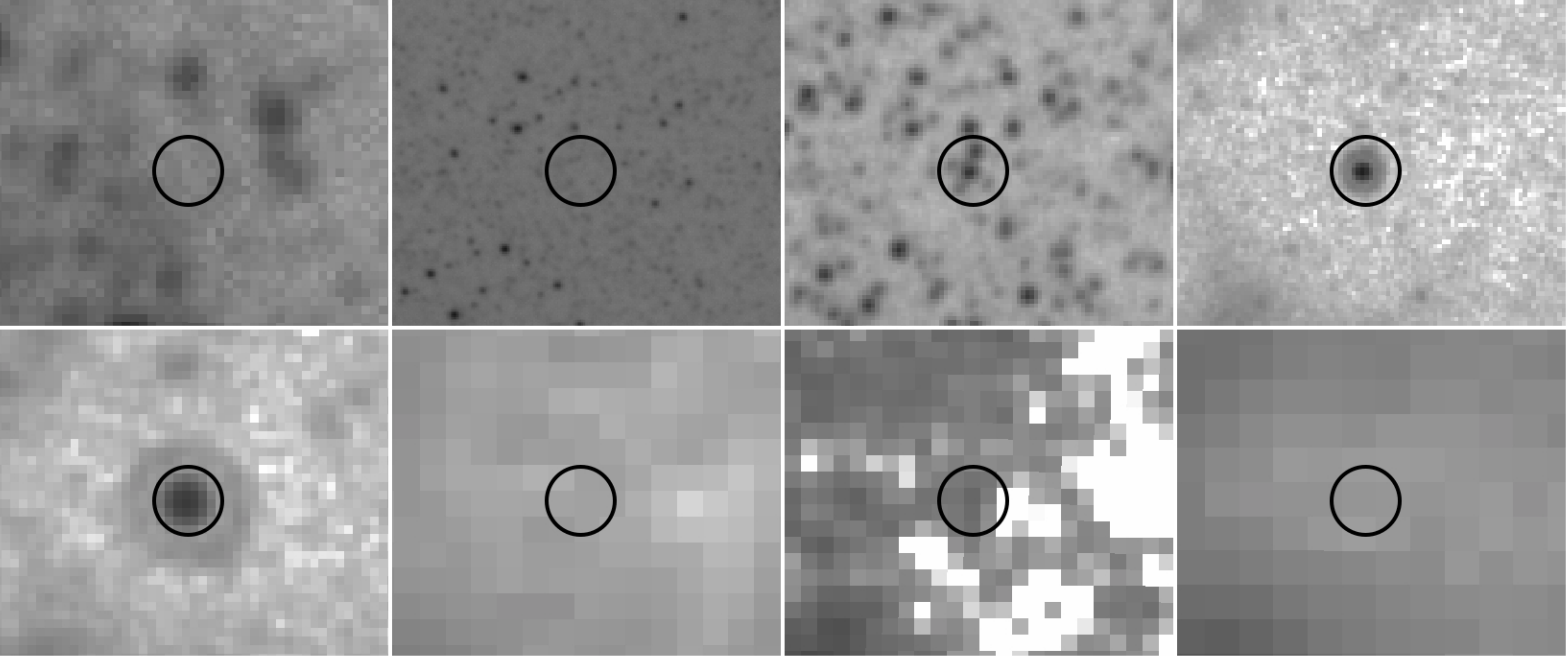}
\vspace*{-5truemm}
\caption{Multi-wavelength postage images of the environment around VC 5 (EAGB-2, Khan et al. 2010). Each image is 1\arcmin\ per side. From left to right the images are: GALEX NUV, LGGS R, IRAC 3.6~\micron, IRAC 8.0~\micron, MIPS 24~\micron, MIPS 70~\micron, PACS 160~\micron, and SPIRE 250~\micron.}\label{fig4}
\end{figure}

\begin{figure}[t]
\epsscale{1.0}
\plotone{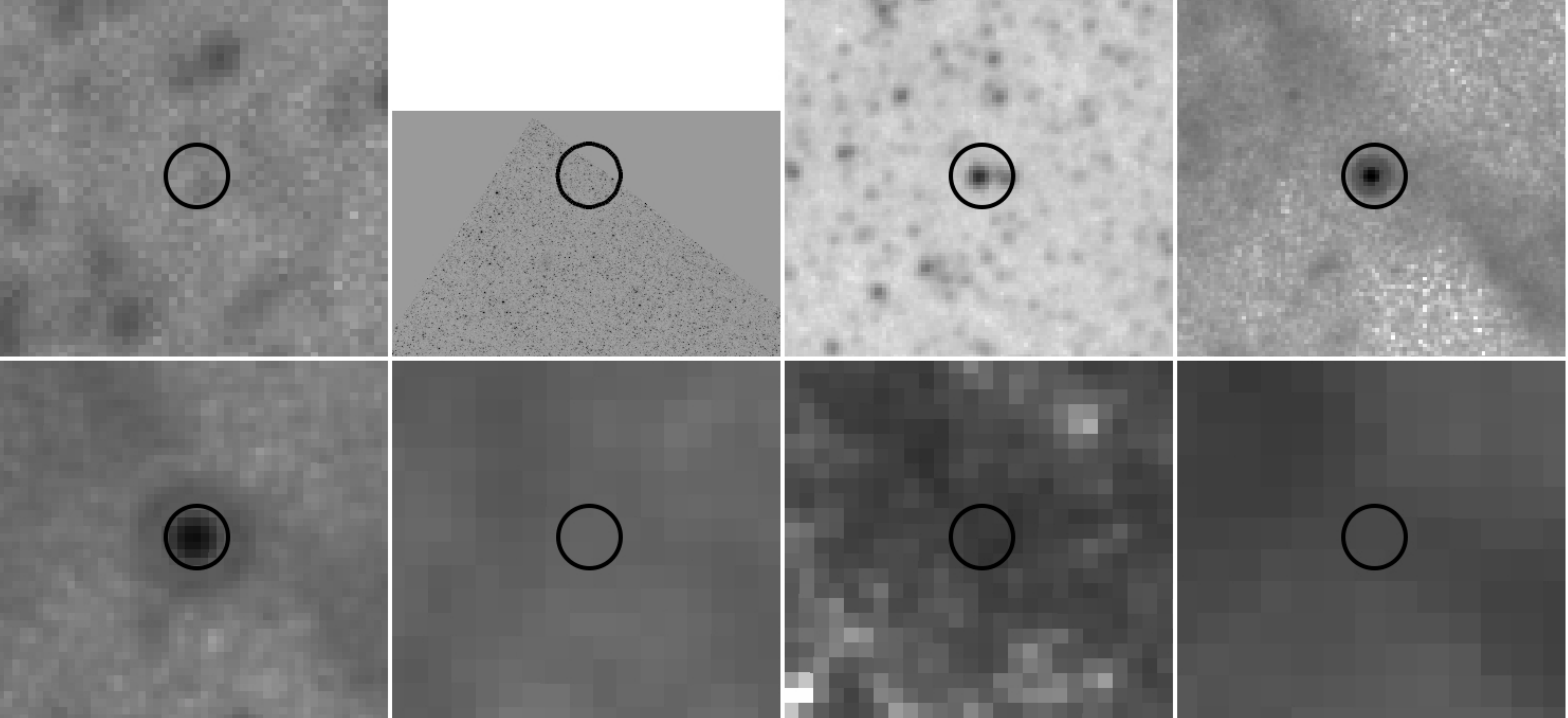}
\vspace*{-5truemm}
\caption{The same as Fig. 4 for VC 7 (SSTM3307 J013412.95+302938.1, Polomski et al. in prep). From left to right the images are: GALEX NUV, HST ACS WFC F606W, IRAC 3.6~\micron, IRAC 8.0~\micron, MIPS 24~\micron, MIPS 70~\micron, PACS 160~\micron, and SPIRE 250~\micron.}\label{fig5}
\end{figure}

\clearpage

\begin{figure}[t]
\epsscale{1.0}
\plotone{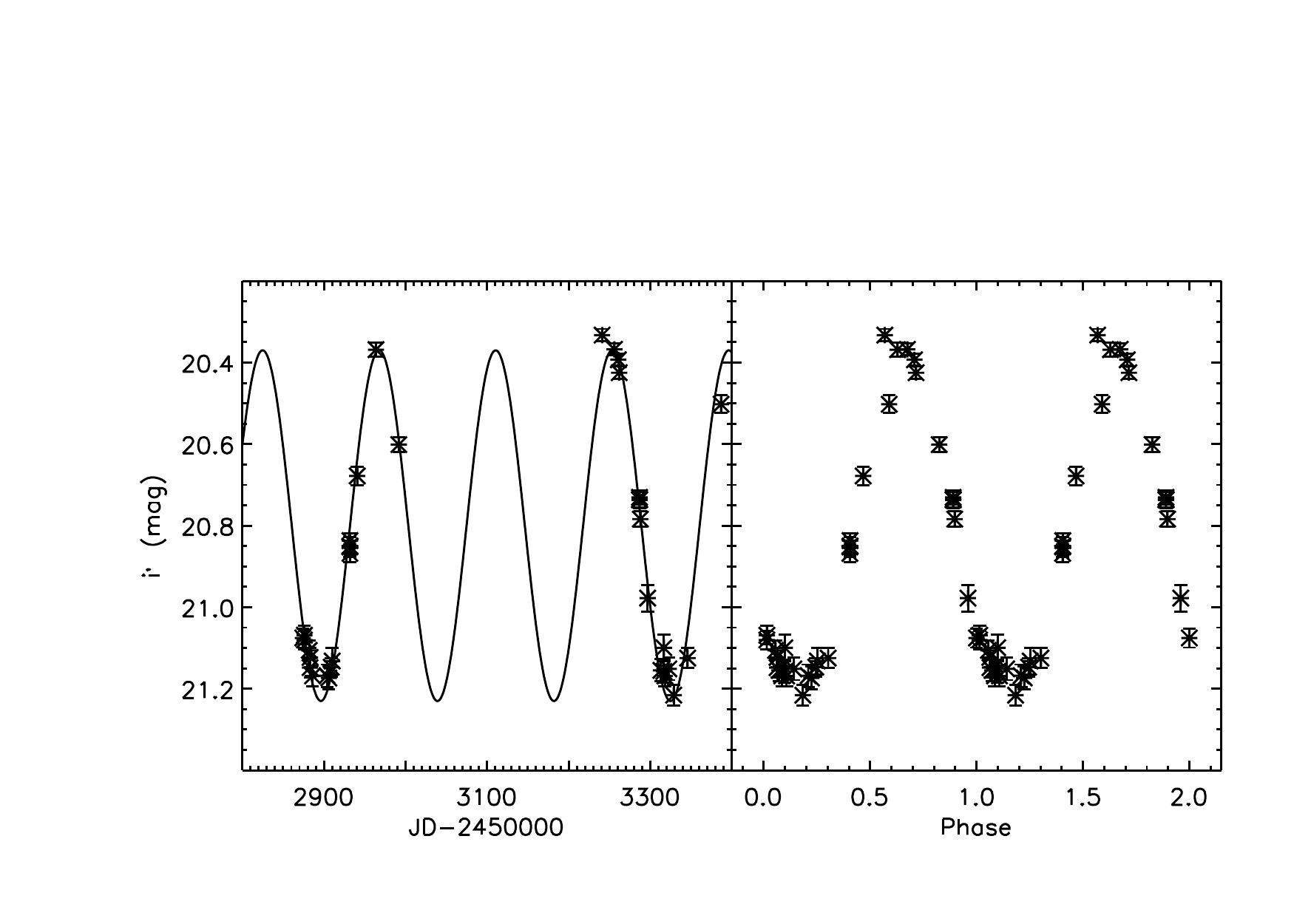}
\vspace*{-10truemm}
\caption{Left: The i$'$ prime light curve for the source from Hartman et al. (2006) that was cross matched to VC 7 with a sine wave fitted to the data (see discussion in text). Right: A phase folded light curve on the period of 142.857 days generated with PERIOD by Starlink.}\label{fig6}
\end{figure}

\clearpage

\begin{figure}[t]
\epsscale{1.0}
\plotone{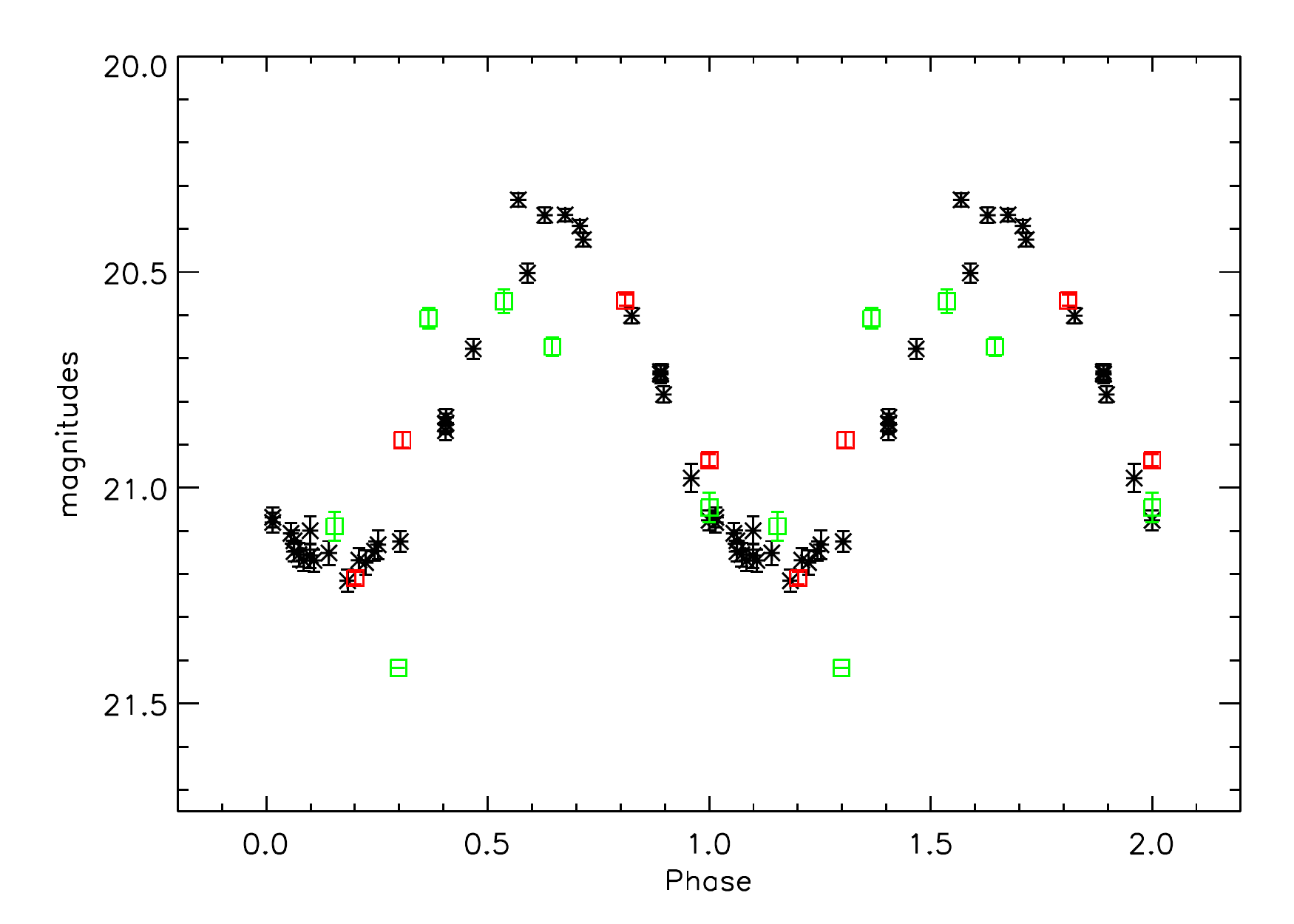}
\vspace*{-10truemm}
\caption{The VC 7 MIPS 24~\micron\ (red squares) and IRAC 8.0~\micron\ (green squares) observations folded on the period determined from the Hartman et al. (2006) data (black asterisks). The MIPS and IRAC data have been shifted so that their mean magnitude is the same as Hartman et al. See \S 5.3 for a discussion on the sixth IRAC epoch.}\label{fig7}
\end{figure}

\clearpage

\begin{figure}[t]
\epsscale{1.0}
\plotone{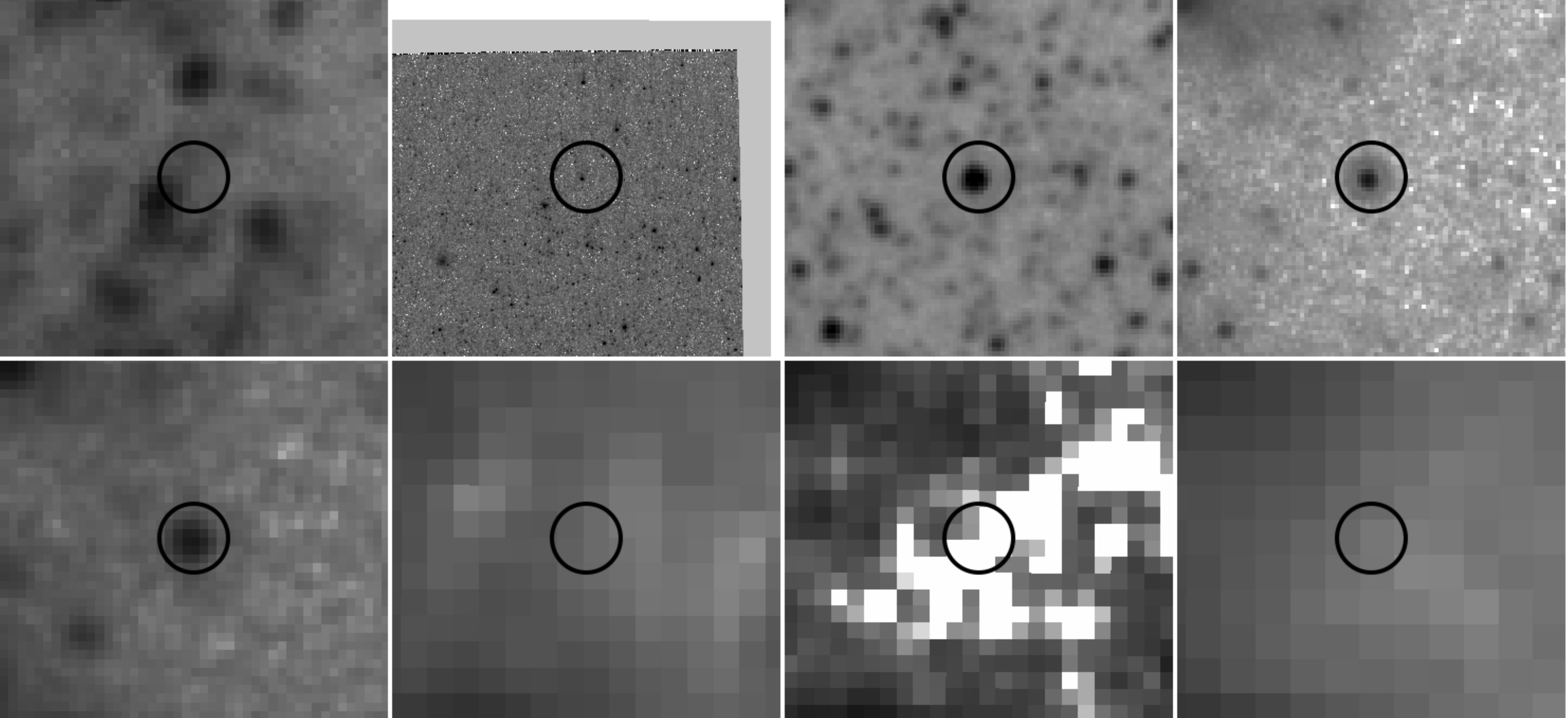}
\vspace*{-5truemm}
\caption{The same as Fig. 4 for VC 14 (VHK 71, van den Bergh et al. 1975). From left to right the images are: GALEX NUV, HST WFPC2 F555W, IRAC 3.6~\micron, IRAC 8.0~\micron, MIPS 24~\micron, MIPS 70~\micron, PACS 160~\micron, and SPIRE 250~\micron.}\label{fig8}
\end{figure}

\clearpage

\begin{figure}[t]
\epsscale{0.95}
\plotone{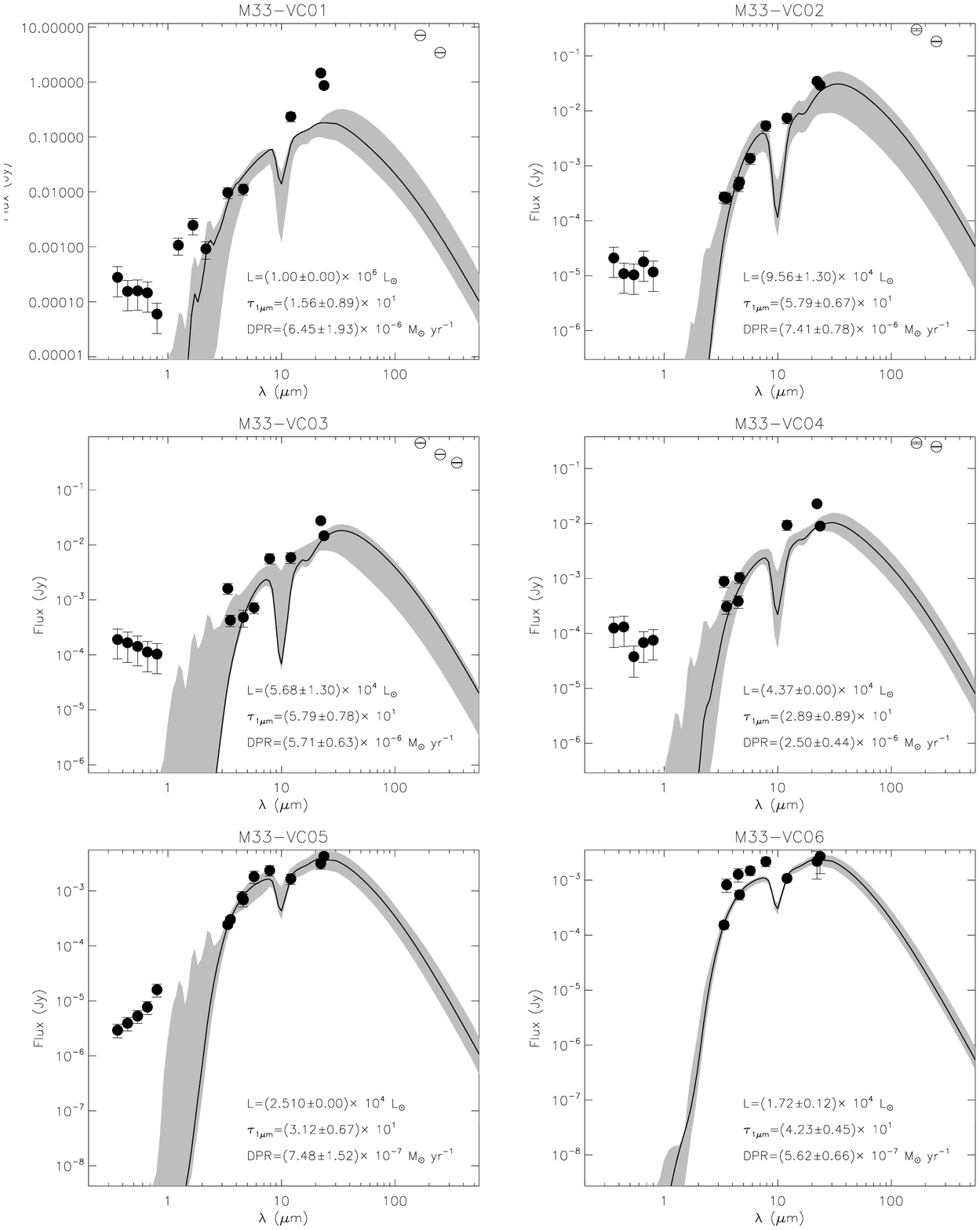}
\vspace*{-10truemm}
\caption{The SEDs with best-fit GRAMS (solid line) model for VCs 1--6. The gray shading is the error on the GRAMS fit, and is determined by the 250 best-fits of the same chemical composition. Filled circles represent LGGS, 2MASS, WISE, IRAC, and MIPS points that were used for the GRAMS fitting. Unfilled circles are additional photometry photometry points at wavelengths longer than 70~\micron\ that were not used in the fitting. The individual best fit GRAMS parameters are also shown. Zeros represent undetermined uncertainties, see \S 6.}\label{fig9}
\end{figure}

\clearpage

\setcounter{figure}{8}

\begin{figure}[t]
\epsscale{0.95}
\plotone{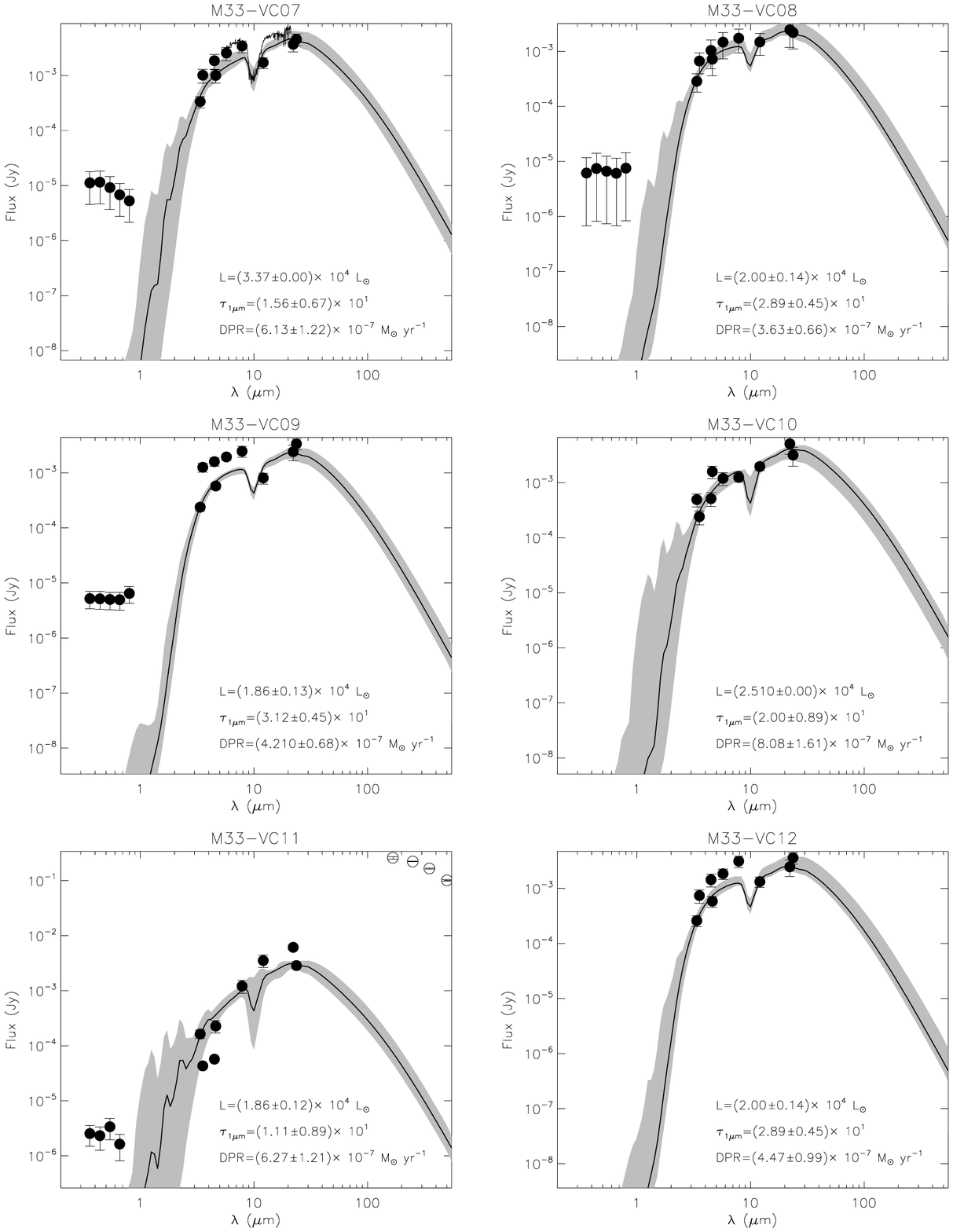}
\vspace*{-10truemm}
\caption{The SEDs with best-fit GRAMS (solid line) model for VCs 7--12. VC 7 includes IRS SL and LL spectra in its best GRAMS fit.}\label{fig9}
\end{figure}

\clearpage

\setcounter{figure}{8}

\begin{figure}[t]
\epsscale{0.95}
\plotone{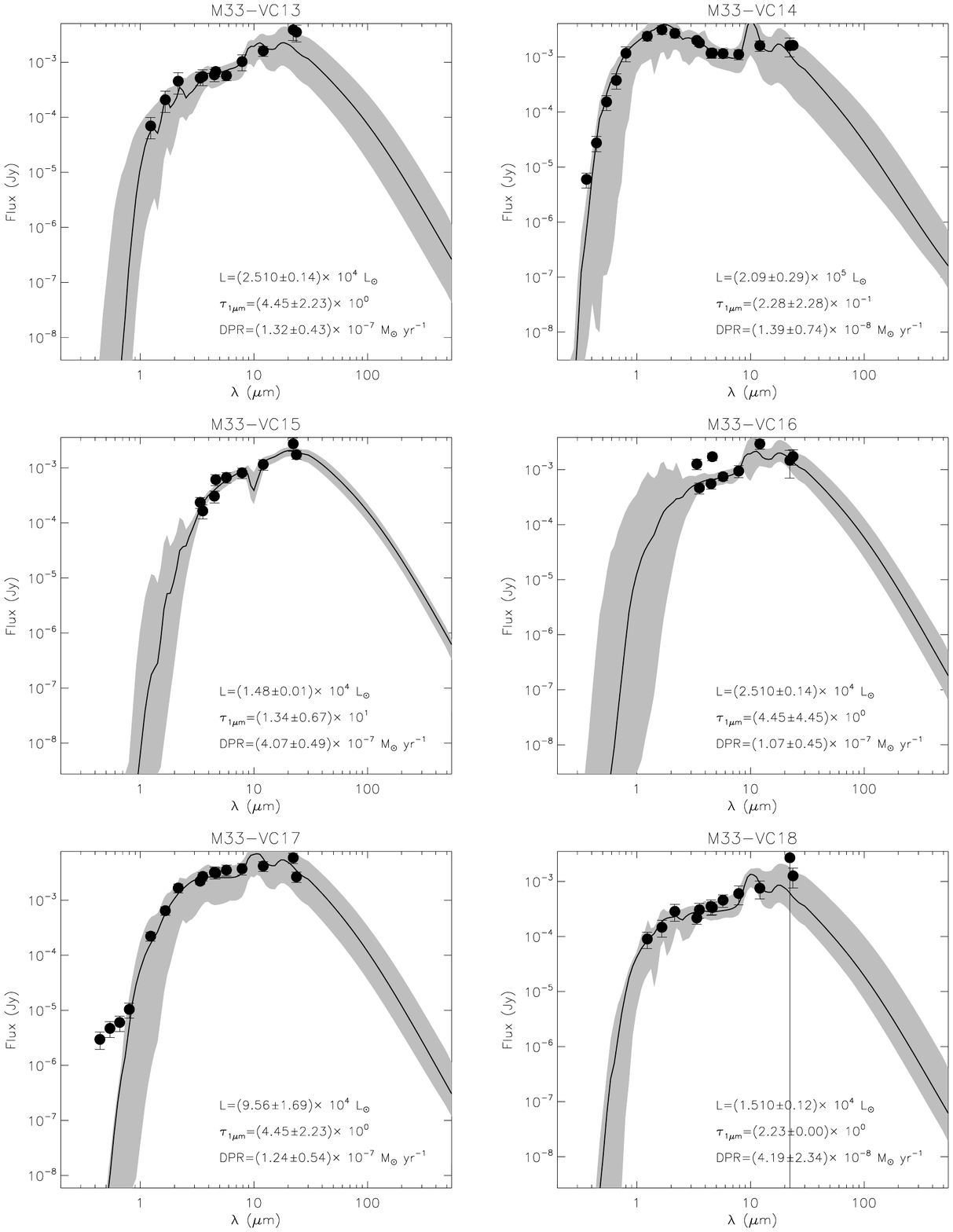}
\vspace*{-10truemm}
\caption{The SEDs with best-fit GRAMS (solid line) model for VCs 13--18. Points with error bars spanning the entire plot are upper limits.}\label{fig9}
\end{figure}

\clearpage

\setcounter{figure}{8}

\begin{figure}[t]
\epsscale{0.95}
\plotone{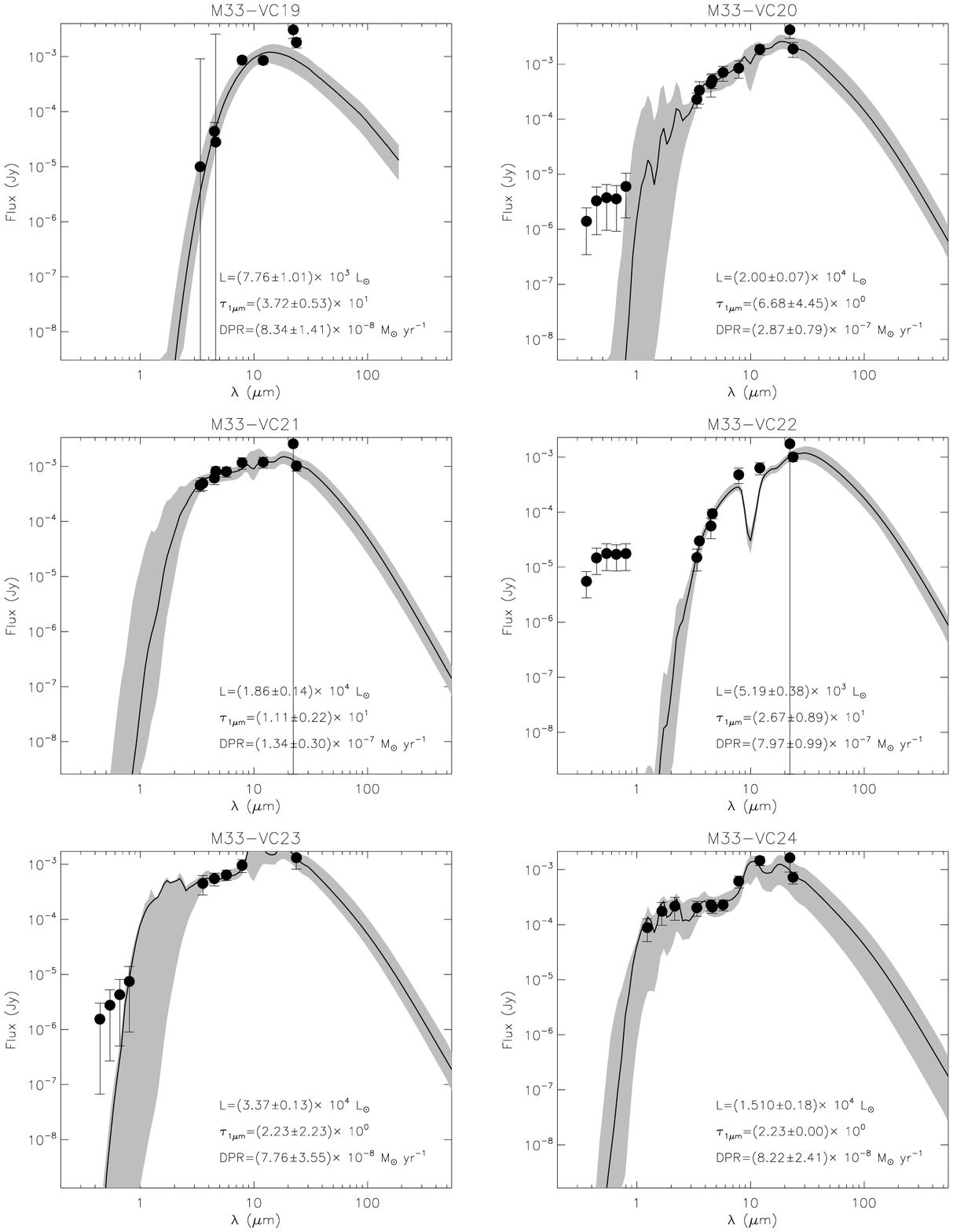}
\vspace*{-10truemm}
\caption{The SEDs with best-fit GRAMS (solid line) model for VCs 19--24. Points with error bars spanning the entire plot are upper limits.}\label{fig9}
\end{figure}

\clearpage 

\begin{figure}[t]
\epsscale{1.0}
\plotone{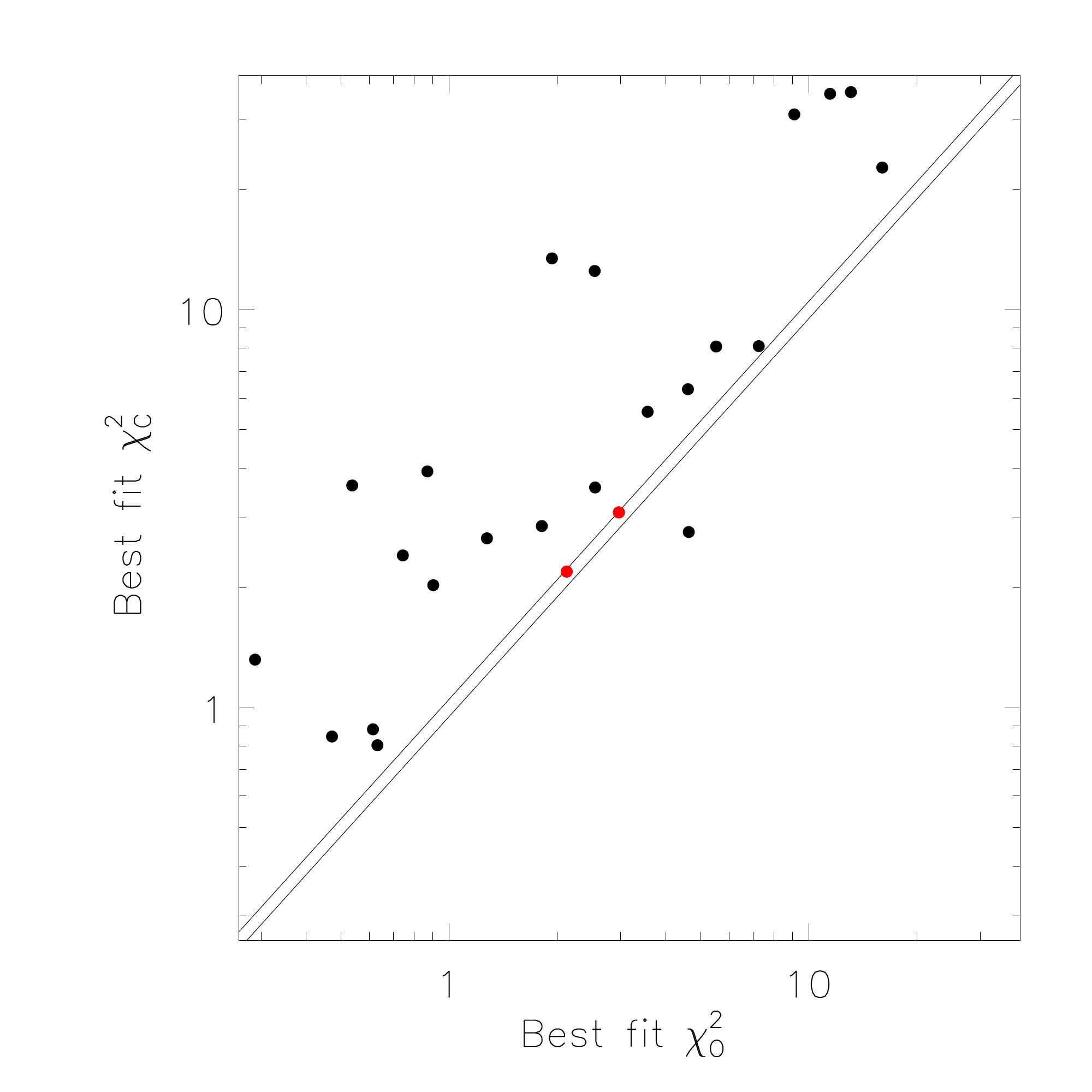}
\vspace*{-10truemm}
\caption{The best fit GRAMS $\chi^2$ for C-rich and O-rich models. Sources above the diagonal lines are C-rich, and below the diagonal lines are O-rich. For the sources between the two lines, which is where the fitting between C-rich and O-rich are within $\pm$ 5\% of each other, have uncertain classification.}\label{fig10}
\end{figure}

\clearpage

\begin{figure}[t]
\epsscale{1.0}
\plotone{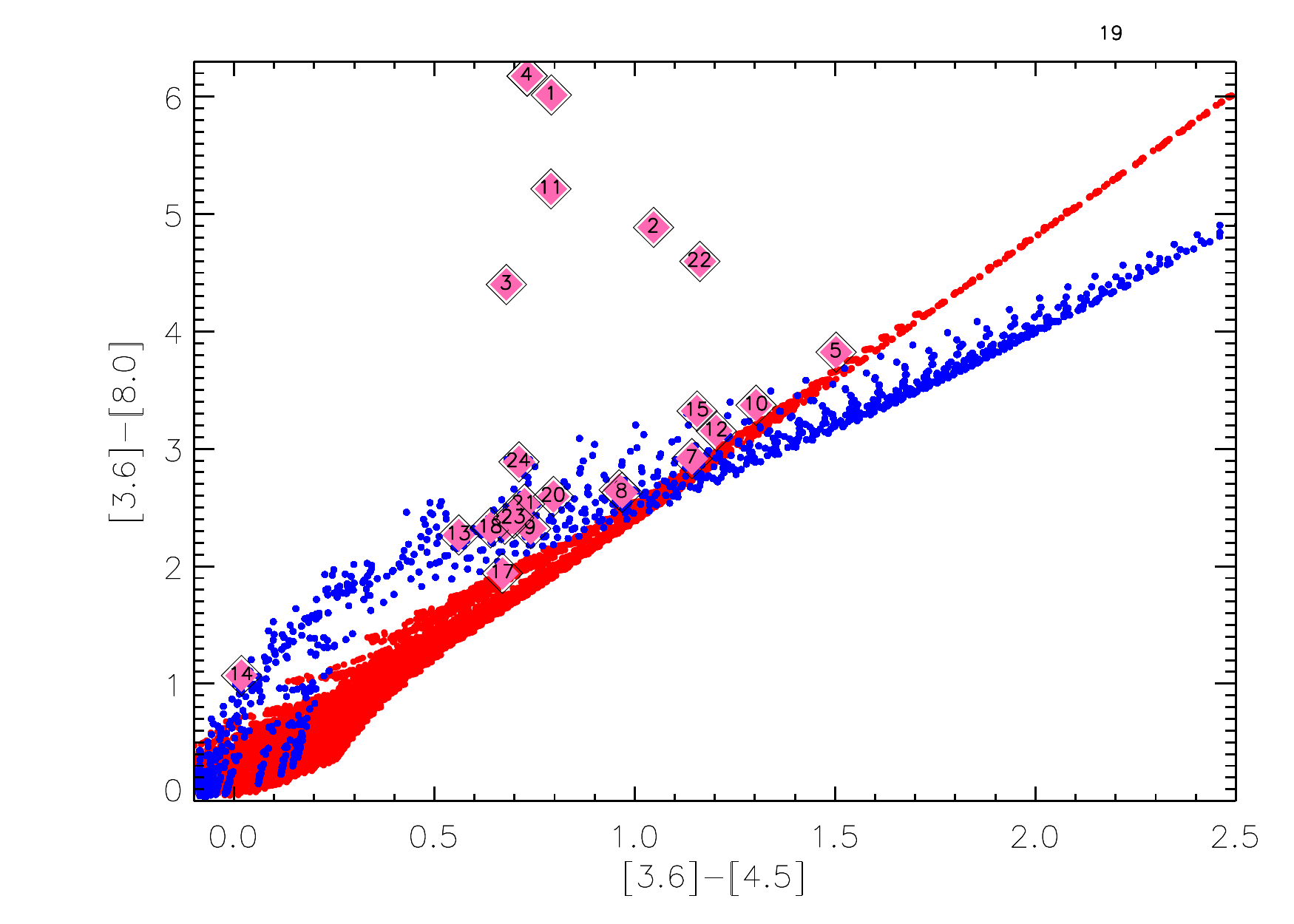}
\vspace*{-10truemm}
\caption{The [3.6]--[8.0] vs [3.6]--[4.5] color--color diagram for our VCs (pink diamonds) overplotted on GRAMS O-rich model grid (blue points) and GRAMS C-rich model grid (red points). IRAC colors alone are not enough to clearly distinguish between the various possible objects that our sources can be.}\label{fig11}
\end{figure}

\clearpage

\begin{figure}[t]
\epsscale{1.0}
\plotone{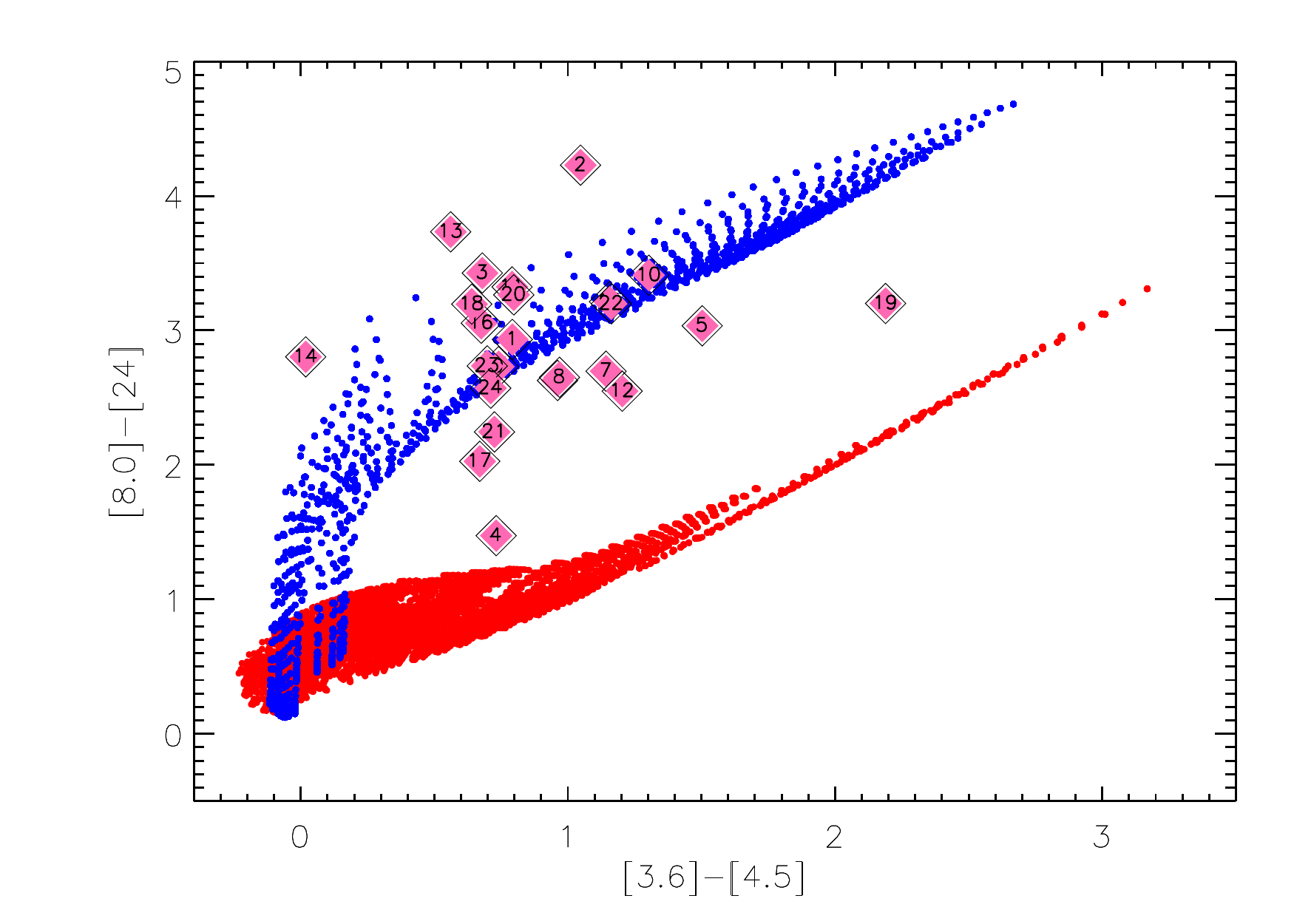}
\vspace*{-10truemm}
\caption{The [8.0]--[24] vs [3.6]--[4.5] color--color diagram for our VCs (pink diamonds) overplotted on GRAMS O-rich model grid (blue points) and GRAMS C-rich model grid (red points). The addition of 24~\micron\ photometry removes the degeneracy between the O-rich and C-rich classifications.}\label{fig12}
\end{figure}

\clearpage

\begin{figure}[t]
\epsscale{1.0}
\plotone{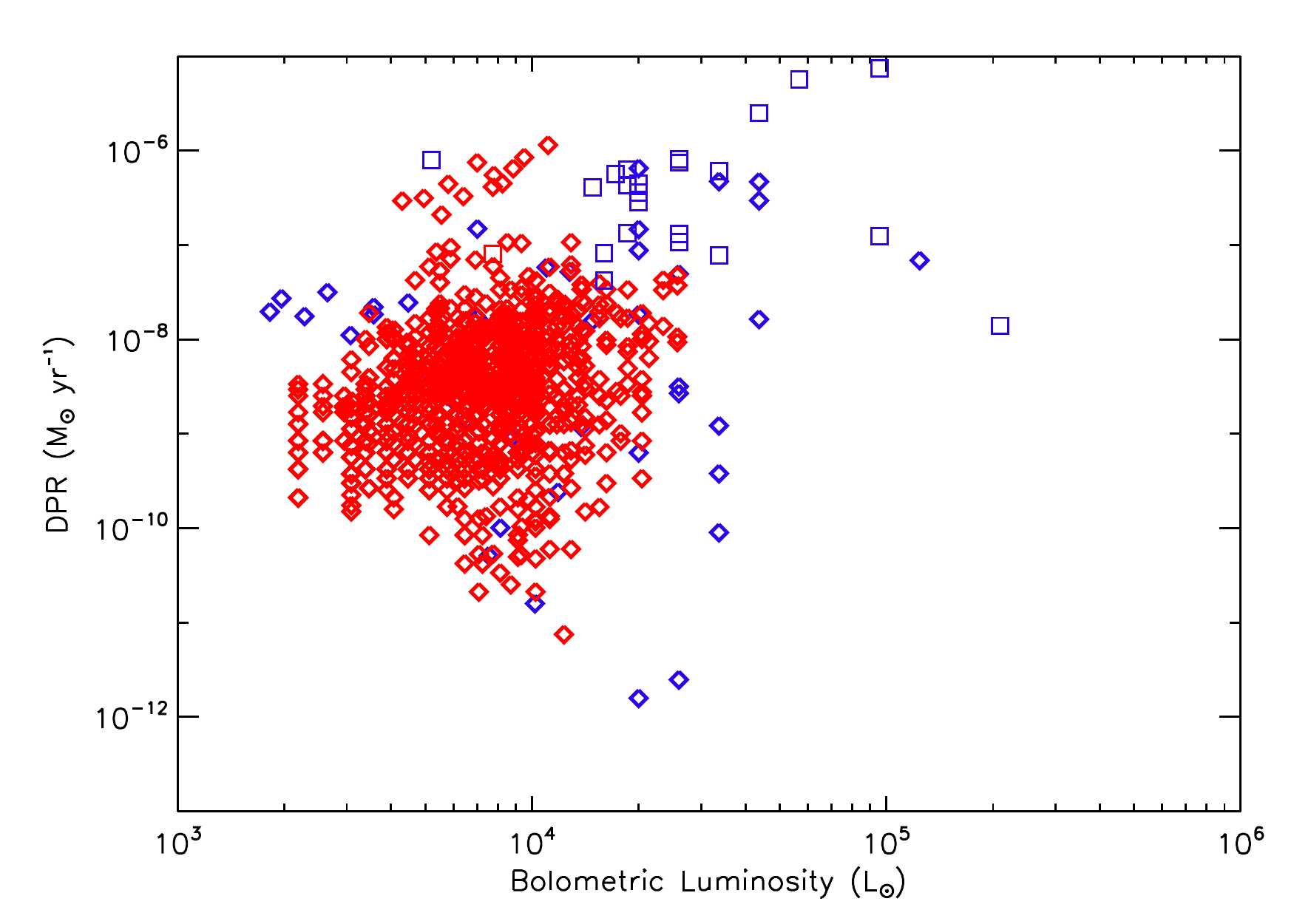}
\vspace*{-10truemm}
\caption{The GRAMS best-fit DPR and bolometric luminosity for our VCs (squares), placed onto Figure 16 from Riebel et al. (2012) with their XAGBs denoted by diamonds. Red and blue represent C-rich and O-rich models, respectively.}\label{fig13}
\end{figure}

\clearpage

\begin{figure}[t]
\epsscale{1.0}
\plotone{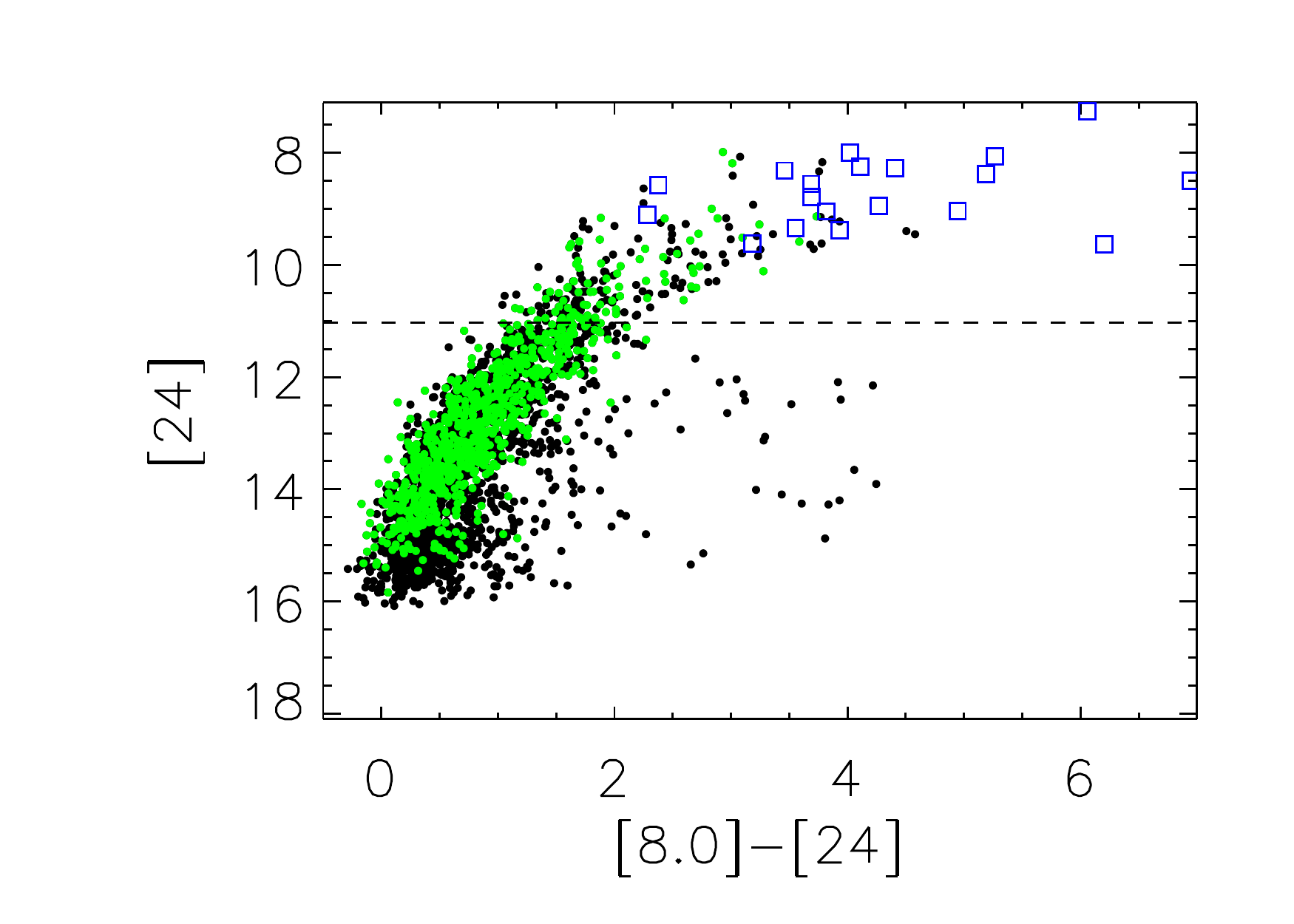}
\vspace*{-10truemm}
\caption{The [24] vs [8.0]--[24.0] CMD from Vijh et al. (2009), where green dots are variable XAGB candidates and black dots are other variable sources. Our VCs (blue squares) are shown to be much redder, and hence more embedded, than the LMC sources. The dashed line represents our 24~\micron\ source PSF detection limit.}\label{fig14}
\end{figure}

\clearpage

\begin{deluxetable}{lcr}
\tablecaption{M33 Available Observations\label{tbl:1}}
\tablewidth{0pt}
\tablehead{
\colhead{Telescope} & 
\colhead{Filter} & 
\colhead{Reference}
}
\startdata
%
          GALEX              &       FUV Imager               &  Gil de Paz et al. 2007              \\
          GALEX              &       NUV Imager              &  Gil de Paz et al. 2007              \\
   Mayall 4.0-m           &    MOSAIC-B band          &  Massey et al. 2006              \\
   Mayall 4.0-m           &  MOSAIC-H$\alpha$      &  Massey et al. 2006              \\
   Mayall 4.0-m           &    MOSAIC-R band          &  Massey et al. 2006              \\
          2MASS             &          2MASS-J                 &  Jarrett et al.2003               \\
          WISE                &  WISE 3.4~\micron\         &  Cutri et al. 2012               \\
  {\it Spitzer}              &  IRAC 3.6~\micron\            &  Dale et al. 2007             \\
  {\it Spitzer}              &  IRAC 4.5~\micron\            &  Dale et al. 2007              \\
         WISE                &  WISE 4.6~\micron\         &  Cutri et al. 2012               \\
  {\it Spitzer}              &  IRAC 5.8~\micron\            &  Dale et al. 2007              \\
  {\it Spitzer}              &  IRAC 8.0~\micron\            &  Dale et al. 2007              \\
         WISE                &  WISE 12~\micron\         &  Cutri et al. 2012               \\
         WISE                &  WISE 22~\micron\         &  Cutri et al. 2012               \\
  {\it Spitzer}              &  MIPS 24~\micron\             &  Dale et al. 2007              \\
  {\it Spitzer}              &  MIPS 70~\micron\             &  Dale et al. 2007              \\
 {\it Herschel}              &  PACS 100~\micron\            &  Kramer et al. 2010              \\
 {\it Herschel}              &  PACS 160~\micron\            &  Kramer et al. 2010              \\
 {\it Herschel}              &  SPIRE 250~\micron\           &  Kramer et al. 2010              \\
 {\it Herschel}              &  SPIRE 350~\micron\           &  Kramer et al. 2010              \\
 {\it Herschel}              &  SPIRE 500~\micron\           &  Kramer et al. 2010              \\

\enddata
\end{deluxetable}

\clearpage

\begin{deluxetable}{lccccccc}
\tablenum{2}
\tablecaption{M33 24~\micron\ Variable Candidates \label{tbl:2}}
\tablewidth{0pt}
\tablehead{
\colhead{VC} & 
\multicolumn{2}{c}{R.A.\hspace*{1em}(J2000)\hspace*{1em} Dec.} &
\colhead{${\langle}F_{24}{\rangle}$} &
\colhead{Amp$_{24}$}&
\colhead{$\chi^2_{PSF}$} &
\colhead{$\chi^2_{Aperture}$}
\\
&
(hh:mm:ss) &
(dd:mm:ss) &
(mJy) &
(mJy) &
&
}
\startdata
 1   &  01:34:32.1  &  +30:47:01  &       864.00   $\pm$        99.80   &  217.30    &      641.0  &       168.0 \\
 2   &  01:34:22.9  &  +30:34:10  &        29.10    $\pm$         2.40  &     4.76 &        37.2  &        19.0 \\
 3   &  01:33:39.1  &  +30:32:36  &        14.70    $\pm$         0.80 &      1.76  &        69.2  &        24.2 \\
 4   &  01:33:32.6  &  +30:36:55  &         8.96   $\pm$         0.46 &      1.10 &       19.5  &        31.2 \\
 5   &  01:34:12.5  &  +30:54:20  &         4.27   $\pm$         0.57 &      1.35 &        56.7  &        29.6 \\
 6   &  01:33:26.8  &  +30:23:13  &         2.72   $\pm$         1.54  &     3.27 &    612.0  &       424.8 \\
 7   &  01:34:12.9  &  +30:29:38  &         4.56   $\pm$         1.13  &     2.72  &   298.5  &       209.6 \\
 8   &  01:33:25.5  &  +30:35:52  &         2.20   $\pm$         1.20  &     2.61  &      381.2  &       190.2 \\
 9   &  01:34:27.8  &  +30:43:40  &         3.38   $\pm$         0.83 &      1.79  &       183.1  &       219.4 \\
10  &  01:33:47.2  &  +30:15:37  &         3.19   $\pm$         1.21  &     2.93 &      308.0  &       247.9 \\
11  &  01:33:03.5  &  +30:39:55  &         2.85   $\pm$          0.14 &     0.33  &      10.6  &        22.9 \\
12  &  01:34:36.1  &  +31:01:35  &         3.61   $\pm$         0.93 &      2.07 &     198.0  &       193.0 \\
13  &  01:33:26.6  &  +30:57:15  &         3.53   $\pm$         1.24  &     2.91 &    555.0  &       393.0 \\
14  &  01:34:12.2  &  +30:53:14  &         1.63   $\pm$         0.22 &      0.43  &    21.5  &        38.7 \\
15  &  01:34:45.7  &  +31:10:14  &         1.73   $\pm$         0.31 &     0.68  &    41.1  &        24.5 \\
16  &  01:33:47.3  &  +30:16:33  &         1.73   $\pm$         0.54 &     1.31  &   122.2  &       109.3 \\
17  &  01:33:49.8  &  +30:52:42  &         2.66   $\pm$         0.63 &     1.38  &    121.7  &       150.0 \\
18  &  01:33:17.5  &  +30:12:28  &         1.26   $\pm$         0.54 &     1.18 &    226.2  &       135.0 \\
19  &  01:32:28.6  &  +30:17:45  &         1.82   $\pm$         0.40 &      0.90 &     83.5  &        47.3 \\
20  &  01:33:19.6  &  +30:31:05  &         1.89   $\pm$         0.59 &     1.38 &   178.4  &        45.3 \\
21  &  01:33:41.5  &  +30:14:13  &         1.02   $\pm$         0.20 &      0.49 &     23.2  &        31.4 \\
22  &  01:33:37.5  &  +30:55:51  &         1.00   $\pm$         0.21 &     0.43 &      32.2  &        19.6 \\
23  &  01:34:09.3  &  +30:55:18  &         1.32   $\pm$         0.52 &     1.21  &     139.0  &        66.5 \\
24  &  01:33:14.6  &  +30:42:27  &         0.73 $\pm$         0.20 &       0.43 &       24.4  &        34.0 \\

\enddata
\end{deluxetable}

\clearpage

\begin{deluxetable}{lcccc}
\tablenum{3}
\tablecaption{M33 24~\micron\ Individual Epoch Photometry\label{tbl:3}}
\tablewidth{0pt}
\tablehead{
\colhead{VC} & 
\colhead{Epoch 1} & 
\colhead{Epoch 2} & 
\colhead{Epoch 3} & 
\colhead{Epoch 4}  
\\
&
(mJy) &
(mJy) &
(mJy) &
(mJy) 
}
\startdata
 1  &       929.00   $\pm$         3.10   &       941.20   $\pm$    3.80   &       724.00 $\pm$        3.40   &     861.00   $\pm$     3.00  \\
 2  &        31.20   $\pm$         0.30  &        31.10   $\pm$       0.34  &        26.50   $\pm$        0.40  &       27.80   $\pm$      0.29 \\
 3  &        15.30   $\pm$         0.10  &        14.90   $\pm$      0.09  &        13.60   $\pm$         0.08  &       14.90   $\pm$      0.09 \\
 4  &         8.43  $\pm$         0.08  &         9.08  $\pm$          0.09  &         8.80  $\pm$           0.08  &        9.53  $\pm$        0.10 \\
 5  &         5.06  $\pm$         0.08  &         4.07  $\pm$         0.07  &         3.71  $\pm$            0.05  &        4.23  $\pm$        0.07 \\
 6  &         4.95  $\pm$         0.07  &         2.48  $\pm$         0.05  &         1.74  $\pm$            0.03  &        1.68  $\pm$        0.04 \\
 7  &         4.32  $\pm$         0.06  &         6.08  $\pm$         0.07  &         4.51  $\pm$            0.07  &        3.36  $\pm$        0.04 \\
 8  &         3.95  $\pm$         0.07  &         2.00  $\pm$         0.05  &         1.51  $\pm$            0.03  &        1.34  $\pm$        0.04 \\
 9  &         3.91  $\pm$         0.06  &         4.22  $\pm$         0.07  &         2.96  $\pm$           0.05  &         2.43  $\pm$        0.05 \\
10  &         3.10 $\pm$         0.06  &         1.88  $\pm$        0.05  &         2.97  $\pm$            0.05  &         4.81  $\pm$        0.07 \\
11  &         2.83 $\pm$         0.03  &         3.00  $\pm$        0.04  &         2.68  $\pm$            0.04  &         2.91  $\pm$        0.02 \\
12  &         2.80 $\pm$        0.05  &         4.87  $\pm$         0.07  &         3.72  $\pm$            0.06  &         3.03  $\pm$        0.04 \\
13  &         1.92 $\pm$        0.04  &         4.82  $\pm$         0.08  &         4.08  $\pm$            0.06  &         3.31  $\pm$        0.05 \\
14  &         1.82 $\pm$        0.04  &         1.39  $\pm$         0.04  &         1.51  $\pm$            0.03  &         1.81  $\pm$        0.04 \\
15  &         1.80 $\pm$        0.04  &         1.28  $\pm$         0.04  &         1.95  $\pm$            0.04  &         1.90  $\pm$        0.04 \\
16  &         1.76 $\pm$        0.04  &         1.12  $\pm$         0.04  &         1.61  $\pm$            0.03  &         2.43  $\pm$        0.05 \\
17  &         1.74 $\pm$        0.05  &         3.11  $\pm$         0.06  &         3.02  $\pm$            0.07  &         2.77  $\pm$        0.05 \\
18  &         1.58 $\pm$        0.03  &         0.66 $\pm$         0.03  &         0.98 $\pm$              0.03  &         1.84  $\pm$        0.03 \\
19  &         1.23 $\pm$        0.04  &         2.01  $\pm$         0.05  &        1.92  $\pm$             0.04  &         2.13  $\pm$        0.05 \\
20  &         1.06 $\pm$        0.04  &         2.11 $\pm$         0.05  &         2.44  $\pm$             0.05  &         1.95  $\pm$        0.04 \\
21  &         1.03 $\pm$        0.04  &         1.25  $\pm$         0.04  &        1.04  $\pm$             0.03  &         0.76 $\pm$         0.04 \\
22  &         0.89 $\pm$       0.03  &         0.76  $\pm$         0.04  &         1.17  $\pm$             0.03  &         1.19  $\pm$        0.03 \\
23  &         0.66 $\pm$       0.03  &         1.87  $\pm$         0.06  &         1.55  $\pm$             0.04  &         1.20  $\pm$        0.04 \\
24  &         0.60 $\pm$       0.03  &         1.00  $\pm$         0.04  &         0.73 $\pm$              0.03  &         0.57 $\pm$         0.03 \\
\enddata
\end{deluxetable}

\begin{deluxetable}{lccccr}
\tablenum{4}
\tablecaption{GRAMS Fitting Results\label{tbl:4}}
\tablewidth{0pt}
\tablehead{
\colhead{VC} & 
\colhead{GRAMS Class} & 
\colhead{L} &
\colhead{DPR} &
\colhead{$\tau_{1\micron}$} &
\colhead{$\chi^2$}
\\
&
&
($10^4$ L$_\odot$)&
($10^{-7}$ M$_\odot~yr^{-1}$) &
&
}
\startdata
 1$^{\rm A}$  & O  & 100.0   &  64.5 $\pm$ 19.3 & 15.6 $\pm$ 8.9 &16.0 \\
 2                  & O   &   9.56  $\pm$  1.30  &  74.1 $\pm$  7.80  & 57.9 $\pm$ 6.7 &    9.10 \\
 3                  & O   &  5.68   $\pm$   1.30  &  57.1 $\pm$ 6.30   &57.9 $\pm$ 7.8 &  11.5 \\
 4                  & O   &   4.37    &  25.0 $\pm$ 4.40   & 28.9 $\pm$ 8.9  & 13.1 \\
 5 		    & O   &   2.51   &  7.48 $\pm$ 1.52   & 31.2 $\pm$ 6.7 & 5.50 \\
 6 		    & U   &   1.72   $\pm$  0.12  &  5.62 $\pm$ 0.66   & 42.3 $\pm$ 4.5 &  3.00$^{\rm B}$\\
 7 	            & O   &   3.37    &  6.13 $\pm$ 1.22   & 15.6 $\pm$ 6.7 &  1.80 \\
 8 	            & O   &   2.00   $\pm$   0.14  &  3.63 $\pm$ 0.66   & 28.9 $\pm$ 4.5 &  0.63 \\
 9 	            & O   &   1.86   $\pm$   0.13  &  4.21 $\pm$ 0.68   & 31.2 $\pm$ 4.5 &  7.30 \\
10 		   & O  &   2.51     &  8.08 $\pm$ 1.70   & 20.0 $\pm$ 8.9 &  1.30 \\
11                & O  &   1.86    $\pm$   0.12  &  6.27 $\pm$ 1.30   & 11.1 $\pm$ 8.9 &  4.60 \\
12 	           & O  &   2.00    $\pm$   0.14  &  4.47 $\pm$ 0.99   & 28.9 $\pm$ 4.5 &  2.50 \\
13 		   & O  &   2.51    $\pm$   0.14  &  1.32 $\pm$ 0.43   & 4.45 $\pm$ 2.23 &  0.74 \\
14 	           & O  & 20.9      $\pm$   2.90  &  0.139 $\pm$ 0.074 & 0.28 $\pm$ 0.28 &  1.93 \\
15 	           & O  &   1.48    $\pm$   0.01  &  4.07 $\pm$ 0.49   & 13.4 $\pm$ 6.7 &  0.87 \\
16 		  & O  &   2.51    $\pm$   0.14  &  1.07 $\pm$ 0.45   & 4.45 $\pm$ 4.45 &  3.60 \\
17 		  & O  &   9.56    $\pm$   1.69  &  1.24 $\pm$ 0.54   & 4.45 $\pm$ 2.23 &  2.50 \\
18 		  & O  &   1.51    $\pm$   0.12  &  0.419 $\pm$ 0.234   & 2.23  &  0.61 \\
19 		  & C  &   0.776  $\pm$   0.101  &   0.834 $\pm$ 0.140  & 37.2 $\pm$ 5.3 &  2.80 \\
20 	          & O  &   2.00    $\pm$   0.07  &  2.87 $\pm$ 0.79   & 6.68 $\pm$ 4.45 & 0.90 \\
21 		  & O  &   1.86    $\pm$   0.14  &  1.34 $\pm$ 0.30   & 11.1 $\pm$ 2.2 &  0.29 \\
22 		  & U  &   0.519  $\pm$   0.038  &  7.97 $\pm$ 0.99   & 26.7 $\pm$ 8.9 &  2.10$^{\rm B}$ \\
23 		  & O  &   3.37    $\pm$   0.13  &  0.776 $\pm$ 0.355   & 2.23 $\pm$ 2.23 &  0.47 \\
24 		  & O  &   1.51    $\pm$   0.18  &  0.822 $\pm$ 0.250   & 2.23  &  0.54 \\
\enddata
\tablecomments{Entries without ``$\pm$'' have undetermined uncertainties, see \S 5.5.1. A: VC1 is co-incident with NGC604. B: VCs 6 \& 22 have uncertain classifications. Their C-rich $\chi^2$  are 3.10 and 2.20, respectively.}
\end{deluxetable}
\end{document}